# Ontologizing Health Systems Data at Scale: Making Translational Discovery a Reality

Tiffany J. Callahan[1,2,*], Adrianne L. Stefanski[1], Jordan M. Wyrwa[3], Chenjie Zeng[4], Anna Ostropolets[2], Juan M. Banda[5], William A. Baumgartner Jr.[1], Richard D. Boyce[6], Elena Casiraghi[7,8], Ben D. Coleman[9], Janine H. Collins[10], Sara J. Deakyne-Davies[11], James A. Feinstein[12], Melissa A. Haendel[13], Asiyah Y. Lin[4], Blake Martin[14], Nicolas A. Matentzoglu[15], Daniella Meeker[16], Justin Reese[8], Jessica Sinclair[17], Sanya B. Taneja[18], Katy E. Trinkley[19], Nicole A. Vasilevsky[20], Andrew Williams[21], Xingman A. Zhang[22], Joshua C. Denny[4], Peter N. Robinson[9], Patrick Ryan[23], George Hripcsak[2], Tellen D. Bennett[14], Lawrence E. Hunter[1,24], Michael G. Kahn[24]

[1]Computational Bioscience Program, University of Colorado Anschutz Medical Campus, Aurora, CO 80045, USA
[2]Department of Biomedical Informatics, Columbia University Irving Medical Center, New York, NY 10032, USA
[3]Department of Physical Medicine and Rehabilitation, School of Medicine, University of Colorado Anschutz Medical Campus, Aurora, CO 80045, USA
[4]National Human Genome Research Institute, National Institutes of Health, Bethesda, MD 20892, USA
[5]Department of Computer Science, Georgia State University, Atlanta, GA 30303, USA
[6]Department of Biomedical Informatics, University of Pittsburgh School of Medicine, Pittsburgh, PA 15260, USA
[7]Computer Science, Università degli Studi di Milano, Milan, 20122, Italy
[8]Division of Environmental Genomics and Systems Biology, Lawrence Berkeley National Laboratory, Berkeley, CA 94720, USA
[9]The Jackson Laboratory for Genomic Medicine, Farmington, CT 06032, USA
[10]Department of Haematology, University of Cambridge, Cambridge, UK
[11]Department of Research Informatics & Data Science, Analytics Resource Center, Children's Hospital Colorado, Aurora, CO 80045, USA
[12]Adult and Child Center for Health Outcomes Research and Delivery Science (ACCORDS), University of Colorado Anschutz School of Medicine, Aurora, CO 80045, USA
[13]Center for Health AI, University of Colorado Anschutz Medical Campus, Aurora CO 80045, USA
[14]Departments of Biomedical Informatics and Pediatrics, University of Colorado School of Medicine, Aurora, CO 80045, USA
[15]Semanticly, Athens, Greece
[16]Yale School of Medicine, New Haven, CT 06510, USA
[17]HealthLinc, Valparaiso, IN 46383, USA
[18]Intelligent Systems Program, University of Pittsburgh, Pittsburgh, PA 15260, USA
[19]Department of Family Medicine, University of Colorado Anschutz School of Medicine, Aurora, CO 80045, USA
[20]Translational and Integrative Sciences Lab, University of Colorado Anschutz Medical Campus, Aurora, CO 80045, USA
[21]Tufts Institute for Clinical Research and Health Policy Studies, Tufts University, Boston, MA 02155, USA
[22]Sema4, Stamford, CT 06902 USA
[23]Janssen Research and Development, Raritan, NJ 08869, USA
[24]Department of Biomedical Informatics, University of Colorado School of Medicine, Aurora, CO 80045, USA

[*]Corresponding author

**ABSTRACT**

**Background:** Common data models solve many challenges of standardizing electronic health record (EHR) data, but are unable to semantically integrate all the resources needed for deep phenotyping. Open Biological and Biomedical Ontology (OBO) Foundry ontologies provide computable representations of biological knowledge and enable the integration of heterogeneous data. However, mapping EHR data to OBO ontologies requires significant manual curation and domain expertise. **Objective:** We introduce OMOP2OBO, an algorithm for mapping Observational Medical Outcomes Partnership (OMOP) vocabularies to OBO ontologies. **Results:** Using OMOP2OBO, we produced mappings for 92,367 conditions, 8611 drug ingredients, and 10,673 measurement results, which covered 68-99% of concepts used in clinical practice when examined across 24 hospitals. When used to phenotype rare disease patients, the mappings helped systematically identify undiagnosed patients who might benefit from genetic testing. **Conclusions:** By aligning OMOP vocabularies to OBO ontologies our algorithm presents new opportunities to advance EHR-based deep phenotyping.

## 1.0 INTRODUCTION

Electronic health record (EHR) adoption, which is nearly universal within the US healthcare system,[1,2] has increased adherence to evidence-based clinical guidelines[3] and facilitated greater patient communication[4] resulting in significant improvements in care.[5] EHRs contain a myriad of systematically collected, longitudinal, patient-level information and are a valuable resource for population-level research.[6] The cornerstone of medicine, diagnosis or clinical phenotyping, aims to identify empirically observable traits exhibited by patients (i.e., signs and symptoms) known to be characteristic of a specific disease.[7] Computational phenotyping is the process of converting clinical phenotypes into computer-executable algorithms in order to identify relevant patients from large sources of clinical data, usually EHRs.[8] One promise of EHR-based computational phenotyping is the ability to perform population-level investigations of mechanistic drivers of disease in diverse patient populations.[9,10] Despite significant progress, this objective remains largely aspirational.[6,11–14]

Traditionally, computational phenotypes have been imprecise due to their exclusive reliance on EHR data, which has been shown to be insufficient at capturing the phenotypic heterogeneity present in most complex diseases.[15–18] Deep phenotyping, or "the precise and comprehensive analysis of phenotypic abnormalities in which the individual components of the phenotype are observed and described",[7] is a fundamental component of precision medicine that requires timely synthesis of multiple types of patient data.[19,20] Deep phenotyping has been successfully applied to rare disease and genetic disorders,[21–33] cancer,[34–40] and pregnancy[41–43] using a variety of clinical and -omic data. Despite large-scale biobanking efforts and resources like the UK Biobank[44] and the All of Us Research Program[45], most EHRs do not systematically integrate nor have the infrastructure to integrate patient-level genomic data or other forms of external knowledge (e.g., scientific literature) with clinical data.[46–48]

Within an EHR, most data used for research (i.e., structured data) are stored using clinical terminologies or vocabularies. Clinical vocabularies are defined as a standard representation of preferred terms which may or may not be hierarchical or have formally defined relationships and are designed to facilitate meaningful and unambiguous information exchange within the medical domain.[49–51] Hundreds of clinical vocabularies have been developed and their use differs by hospital and country. Examples include the International Classification of Diseases (ICD),[52] the Logical Observation Identifiers, Names and Codes (LOINC),[53] the Systematized Nomenclature of Medicine -- Clinical Terms (SNOMED-CT),[54] and RxNorm.[55] Most clinical vocabularies were not designed to be integrated or interoperable with other vocabularies, which is one of the long standing barriers preventing the secondary use of EHR data for research.[48] Common data models (CDMs) like the Observational Medical Outcomes Partnership (OMOP)[56] have solved many of the challenges of standardizing, representing, and utilizing clinical EHR data. Unfortunately, most CDMs and associated terminology management systems are not yet able to integrate and interpret genomic data or other sources of external knowledge or publicly available data.[57]

Similar to clinical vocabularies, ontologies are classification systems that provide detailed representations of our knowledge of a specific domain.[51] Ontologies, like those in the Open Biological and Biomedical Ontology (OBO) Foundry, exist for nearly all scales of biological organization and when combined, can provide a semantically rich and biologically accurate representation of molecular entities and mechanisms.[58–60] Unlike clinical vocabularies, ontologies are semantically computable and interoperable with formally defined relationships, which means they can be logically verified and integrated with data from basic science and clinical research.[51] Mapping clinical vocabularies to ontologies has been recognized as a fundamental requirement for use in deep phenotyping.[20,48,51,61] An example of how aligning these resources improves deep phenotyping was recently demonstrated by Zhang et al., (2019)[62] who mapped LOINC[53] to the Human Phenotype Ontology (HPO),[63] which enabled the harmonization of laboratory tests with different clinical codes to common HPO concepts.

Due to the time-consuming manual effort required to map clinical vocabularies to OBO Foundry ontologies, no comprehensive mapping across commonly used ontologies currently exists. While automated mapping approaches exist, they largely remain unable to correctly capture the complex semantics underlying clinical data and the knowledge encoded by clinical vocabulary concepts.[64] For example, when mapping the concept *Peptic Ulcer without Hemorrhage AND without Perforation but with Obstruction* (SNOMED-CT:54157007) to the HPO, most automated approaches would return a single best mapping, most likely *Peptic Ulcer* (HP:0004398). This HPO concept is much broader in meaning than the clinical concept. A more precise mapping would explicitly capture the presence and absence of all relevant phenotypic features: *Peptic Ulcer* (HP:0004398) and *Gastrointestinal Obstruction* (HP:0004796) and NOT *Gastrointestinal Hemorrhage* (HP:0002239) or *Intestinal Perforation* (HP:0031368). To the best of our knowledge no existing mappings or mapping algorithms are capable of capturing this type of complex semantics.

Building on LOINC2HPO, the goal of this work is to develop OMOP2OBO, an algorithm that enables semantically interoperable mappings between clinical vocabularies in the OMOP CDM to OBO Foundry ontologies (Figure 1). The resulting mappings will enhance the semantic interoperability of the data represented by the OMOP concepts and have the potential to advance deep EHR-based phenotyping by enabling the identification of relevant patients using our knowledge of the molecular mechanisms underlying disease rather than billing codes which are prone to error and subject to bias. Using OMOP2OBO, we created the first healthcare system-scale mappings between clinical vocabularies in the OMOP CDM and eight of the most widely used OBO Foundry ontologies[59] spanning diseases, phenotypes, anatomical entities, organisms, chemicals, vaccines, and proteins. The mappings were evaluated on: (I) accuracy, examined by a team of domain experts; (II) generalizability, examined through comparison to a large set of mapped concepts used at least once in clinical practice from 24 hospital systems; and (III) clinical utility, examined through the identification of patients with an undiagnosed rare disease. OMOP2OBO is open source (https://github.com/callahantiff/OMOP2OBO) and includes a custom built interactive dashboard (http://tiffanycallahan.com/OMOP2OBO_Dashboard/).

## 2.0 RESULTS

Supplementary Table 1 lists the acronyms and definitions used in the paper. The resources used to build and evaluate the OMOP2OBO algorithm and mappings are described in Supplementary Table 2 and Supplementary Table 3.

### 2.1 OMOP2OBO Mapping Data

#### 2.1.1 OMOP Data

The OMOP2OBO mappings were created using a de-identified pediatric dataset from the Children's Hospital of Colorado (CHCO) normalized to the OMOP CDM (referred throughout the manuscript as "CHCO OMOP Database" and described in detail in Supplemental Table 3).[56,65] Standardized vocabularies are a fundamental component of the OMOP CDM, which serve as primary vocabularies within each OMOP domain; all other vocabularies within a specific domain are aligned to a standard vocabulary using mappings provided by the CDM. The standard vocabularies used in this work included: SNOMED-CT[54] (the OMOP Condition domain for diseases and clinical findings), RxNorm[55] (the OMOP Drug domain for drug products and vaccines), and LOINC[53] (the OMOP Measurement domain for laboratory tests and assessment scales).[66] Concepts from these three vocabularies, including labels, synonyms, source codes (i.e., standard vocabulary codes), and ancestor concepts obtained from the OMOP CDM, were extracted and used as input to the OMOP2OBO mapping algorithm. Using the CHCO OMOP Database, concepts were organized into two data waves according to whether or not they had been used at least once in clinical practice (i.e., *Concepts Used in Practice*) or not (i.e., *Concepts Not Used in Practice*). Only *Concepts Used in Practice* were manually mapped.

The counts of concepts eligible for mapping by OMOP domain and data wave are shown in Table 1. There were 109,719 condition concepts (*Concepts Used in Practice:* n=29,129,

**Figure 1: Overview of the OMOP2OBO Algorithm.**

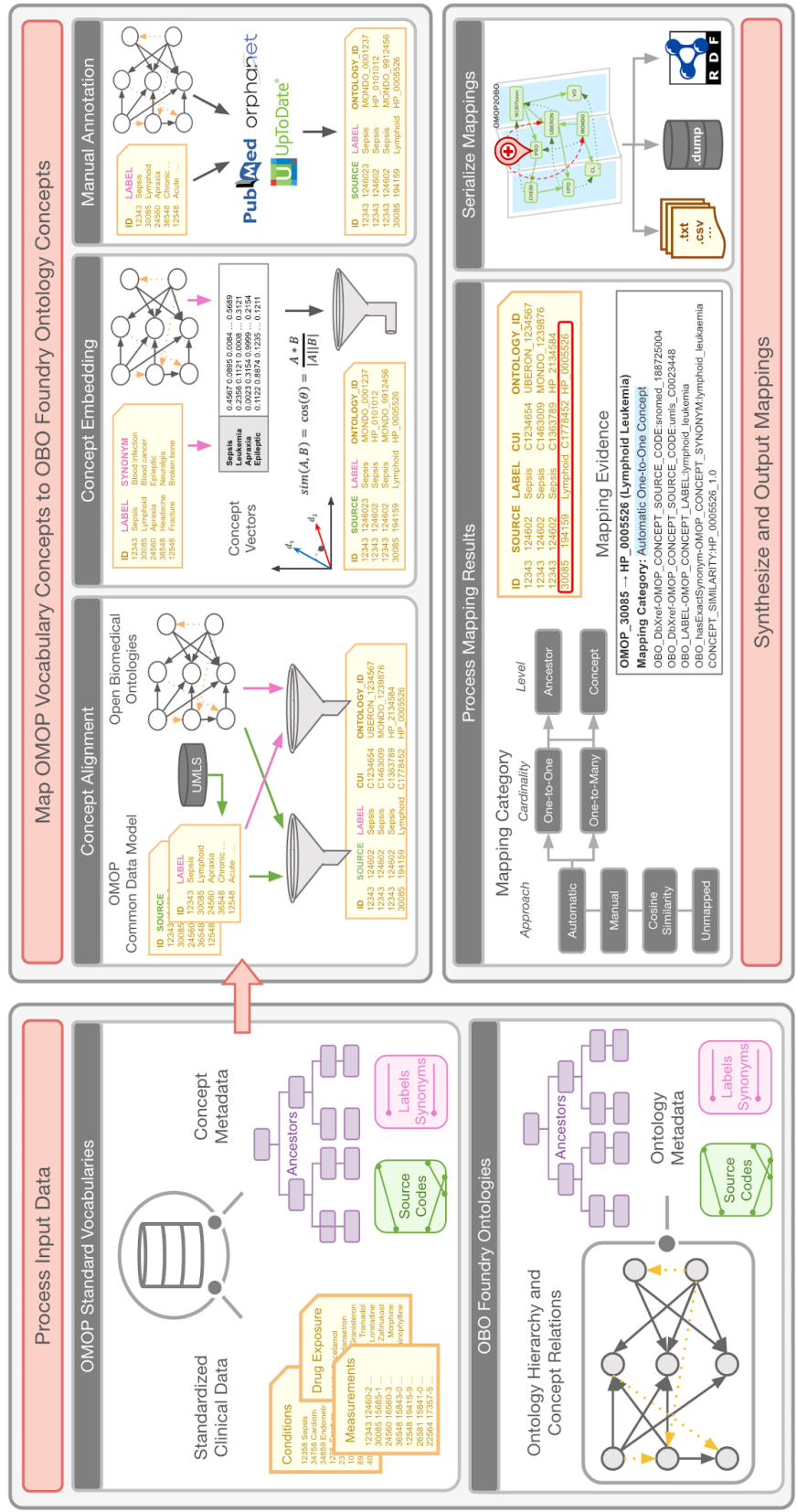

*Concepts Not Used in Practice*: n=80,590) and 11,803 drug ingredient concepts (*Concepts Used in Practice:* n=1693, *Concepts Not Used in Practice*: n=10,110) available to map. For measurements, there were 4083 concepts, representing 11,269 measurement results (*Concepts Used in Practice:* n=1606 concepts [4425 results], *Concepts Not Used in Practice:* n=2477 concepts [6844 results]) available to map. With respect to the *Concepts Used in Practice*, the 29,129 conditions had a median frequency of 25 (range 1-544,618), the 1693 drug ingredients had a median frequency of 251 (range 1-2,267,866), and the 1606 measurement concepts had a median frequency of 313.5 (range 1-56,823,139).

### 2.1.2 OBO Foundry Ontologies

Under the guidance of domain experts, eight OBO Foundry ontologies were selected to represent the following domains: diseases (Mondo[67]), phenotypes (HPO[63]), anatomical entities (Uber Anatomy Ontology [Uberon][68]; Cell Ontology [CL][69]), organisms (National Center for Biotechnology Information Taxon Ontology [NCBITaxon][70]), chemicals (Chemical Entities of Biological Interest [ChEBI][71]), vaccines (the Vaccine Ontology [VO][72]), and proteins (the Protein Ontology [PRO][73]). Each set of ontology concepts also included metadata, which was obtained by querying each ontology for labels, definitions, synonyms, and database cross-references (i.e., codes from other vocabularies and ontologies). The amount of metadata available for mapping is shown in Table 1 and varied across the OBO Foundry ontologies, with NCBITaxon containing the most metadata and Uberon containing the least (visualized in Supplementary Figure 1). A Chi-squared test of independence with Yate's correction revealed a significant association between the ontology and the amount of available metadata ($\chi^2(14)$ = 2,664,853.8, p<0.0001). Post-hoc tests with Bonferroni adjustment confirmed the ontologies provided significantly different amounts of metadata (p<0.0001 for all significant comparisons).

### 2.2 OMOP2OBO Mappings

Figure 2 includes example mappings and illustrates how the OBO Foundry ontologies were used to map concepts from each OMOP domain. As illustrated by this figure, OMOP conditions were mapped to HPO and Mondo, OMOP drug ingredients were mapped to ChEBI, NCBITaxon, PRO, and VO, and OMOP measurements results were mapped to HPO, Uberon, NCBITaxon, PRO, CheBI, and CL. As illustrated in the bottom panel of Figure 1, each mapping consists of four elements: (I) the approach used to create it (i.e., "automatic", "manual", or "cosine similarity"); (II) cardinality (i.e., one-to-one or one-to-many); (III) level (i.e., concept or ancestor); and (IV) evidence, which consists of pipe-delimited free-text phrases that explain what fields were used to construct the mapping. Supplementary Table 4 provides additional details on and examples of the OMOP2OBO mapping categories. The mapping procedures and resources are described in the **OMOP2OBO Algorithm** section of the Methods.

**Figure 2: OMOP2OBO Mapping Examples by OMOP Domain.**

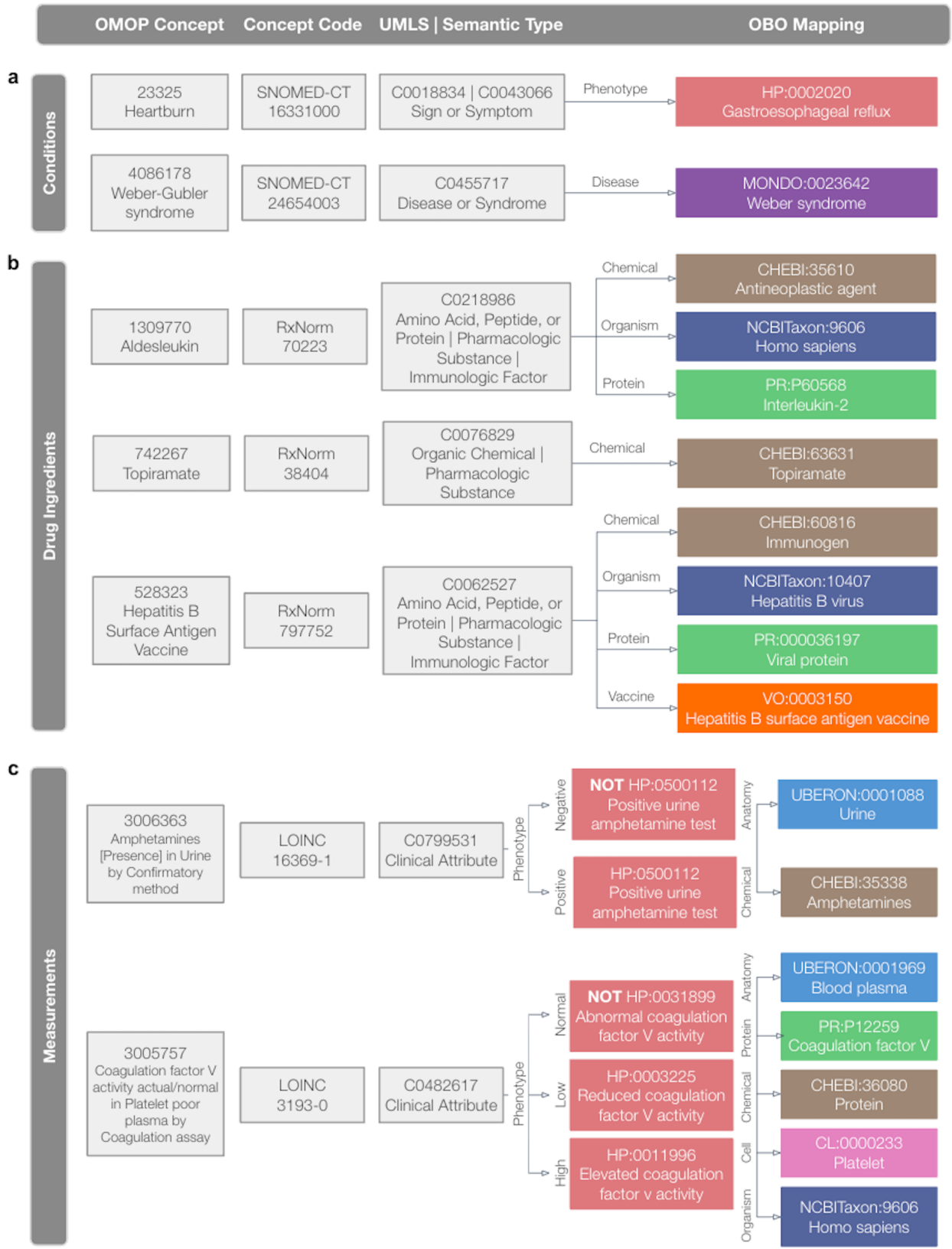

### *2.2.1 Conditions*

Unified Medical Language System (UMLS)[74] concept unique identifiers (CUIs) were found for 96.6% of condition concepts (n=105,976) representing 69 unique Semantic Types.[75] The mapping results for each OBO Foundry ontology are displayed in Figure 3 and detailed in Supplementary Table 5. Of the 109,719 available concepts, 66.9% (n=73,417) mapped to 5654 unique HPO concepts (*Concepts Used in Practice*: 83.9%, *Concepts Not Used in Practice*: 60.8%) and 57.8% (n=63,374) mapped to 9637 unique Mondo concepts (*Concepts Used in Practice*: 68.9%, *Concepts Not Used in Practice*: 53.8%). Only 50 concepts we attempted to map (excluding purposefully unmapped concepts) were unable to be mapped to at least one OBO Foundry ontology concept.

**Mapping Categories.** The frequency distributions of the *Concepts Used in Practice* by mapping category and ontology are visualized in Figure 4. The majority of automatic mappings were one-to-one at the concept-level for *Concepts Used in Practice* (HPO: n=3601, Mondo: n=4836) and *Concepts Not Used in Practice* (HPO: n=1166, Mondo: n=4261). The majority of the manual mappings were one-to-many (HPO: n=10,328, Mondo: n=2835). Cosine similarity-scored concept embeddings enabled 1374 HPO (*Concepts Used in Practice*: median 0.5, range 0.2-1; *Concepts Not Used in Practice:* median 0.4, range 0.2-1) and 667 Mondo (*Concepts Used in Practice*: median 0.8, range 0.2-1; *Concepts Not Used in Practice*: median 1, range 0.2-1) mappings (Supplementary Figure 2a). On average, more evidence was found for mappings to *Concepts Not Used in Practice* than *Concepts Used in Practice* for HPO (8.9 vs 3.9) and Mondo (12.4 vs 10.6).

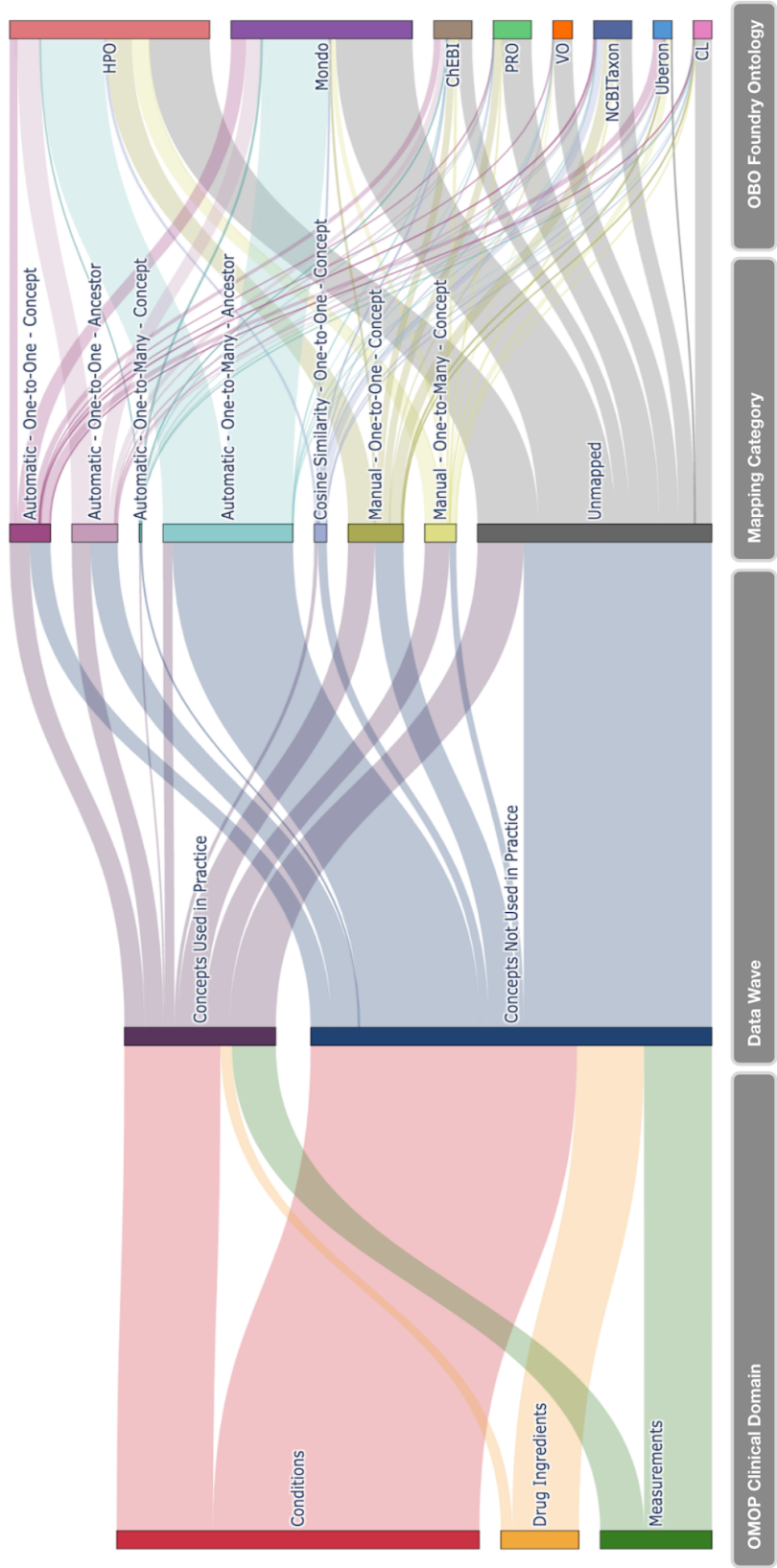

Figure 3: OMOP Concept Mapping Results by Clinical Domain, Concept Type, Mapping Category, and OBO Foundry Ontology.

**Figure 4: Condition Concept Frequency of Use in Clinical Practice by Mapping Category and OBO Foundry Ontology.**

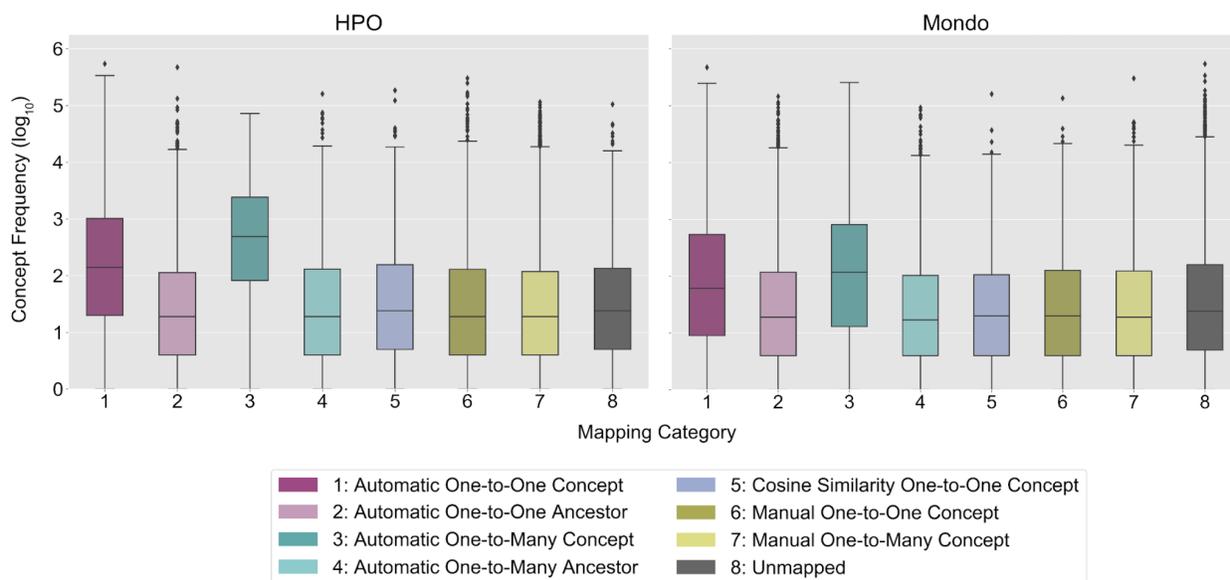

### 2.2.2 Drug Ingredients

UMLS CUIs were found for 99.3% of drug ingredient concepts (n=11,716) representing 23 unique Semantic Types. The mapping results for each OBO Foundry ontology are displayed in Figure 3 and detailed in Supplementary Table 6. Of the 11,803 available concepts, 37.4% (n=411) mapped to 4072 unique ChEBI concepts (*Concepts Used in Practice*: 100%, *Concepts Not Used in Practice*: 26.9%), 21.5% (n=4661) mapped to 2535 unique NCBITaxon concepts (*Concepts Used in Practice*: 23.9%, *Concepts Not Used in Practice*: 42.1%), 2.11% (n=4249) mapped to 142 unique PRO concepts (*Concepts Used in Practice*: 10.5%, *Concepts Not Used in Practice*: 0.7%), and 1.3% (n=154) mapped to 132 unique VO concepts (*Concepts Use in*

*Clinical Practice*: 6.9%, *Concepts Not Used in Practice*: 0.4%). All of the OMOP concepts were mapped to at least one ChEBI concept.

**Mapping Categories.** The frequency distributions of the *Concepts Used in Practice* by mapping category and OBO Foundry ontology are visualized in Figure 5. The majority of automated mappings were one-to-one at the concept-level for *Concepts Used in Practice* (ChEBI: n=959, NCBITaxon: n=20, PRO: n=1, VO: n=90) and *Concepts Not Used in Practice* (ChEBI: n=2192, NCBITaxon: n=135, PRO: n=42, VO: n=18). The majority of the manual mappings were one-to-one (321 ChEBI: n=321, NCBITaxon: n=230, PRO: n=157, VO: n=21). Cosine similarity-scored concept embeddings enabled 109 ChEBI (*Concepts Used in Practice*: median 1, range 0.3-1; *Concepts Not Used in Practice:* median 1, range 0.3-1), 4241 NCBITaxon (*Concepts Used in Practice*: median 0.6, range 0.3-1; *Concepts Not Used in Practice:* median 0.6, range 0.3-1), 18 PRO (*Concepts Used in Practice*: median 0.8, range 0.4-1; *Concepts Not Used in Practice:* median 1, range 0.6-1), and 17 VO (*Concepts Used in Practice*: median 1, range 0.4-1; *Concepts Not Used in Practice:* median 0.8, range 0.4-1) mappings (Supplementary Figure 2b). On average, more evidence was found for mappings to *Concepts Not Used in Practice* than *Concepts Used in Practice* for ChEBI and PRO, excluding NCBITaxon and VO (ChEBI: 7.6 vs 7.6; PRO: 3.9 vs. 1; NCBITaxon: 1.2 vs 1.1; VO: 3 vs. 4.1).

*2.2.3 Measurements*

UMLS CUIs were found for 94.8% of measurement concepts (n=3869) representing a single Semantic Type. The mapping results for each OBO Foundry ontology are displayed in Figure 3 and detailed in Supplementary Table 7. Of the 11,269 measurement results, 96.6% (n=10,888) mapped to 1115 unique HPO concepts (*Concepts Used in Practice*: 92.4%, *Concepts Not Used in Practice*: 99.4%) and 45 unique Uberon concepts (*Concepts Used in Practice*: 92.4%, *Concepts Not Used in Practice*: 99.4%), 76.8% (n=8657) mapped to 425 unique NCBITaxon concepts (*Concepts Used in Practice*: 64.4%, *Concepts Not Used in Practice*: 84.9%), 42.6% (n=4804) mapped to 172 unique PRO concepts (*Concepts Used in Practice*: 35.5%, *Concepts Not Used in Practice*: 47.2%), 87.9% (n=9,904) mapped to 443 unique ChEBI concepts (*Concepts Used in Practice*: 78.9%, *Concepts Not Used in Practice*: 93.7%), and 9.9% (n=1114) mapped to 38 unique CL concepts (*Concepts Used in Practice*: 15.3%, *Concepts Not Used in Practice*: 6.4%). Only 13 concepts we attempted to map (excluding purposefully unmapped concepts) were unable to be mapped to at least one OBO Foundry ontology concept.

**Figure 5: Drug Ingredient Concept Frequency of Use in Clinical Practice by Mapping Category and OBO Foundry Ontology.**

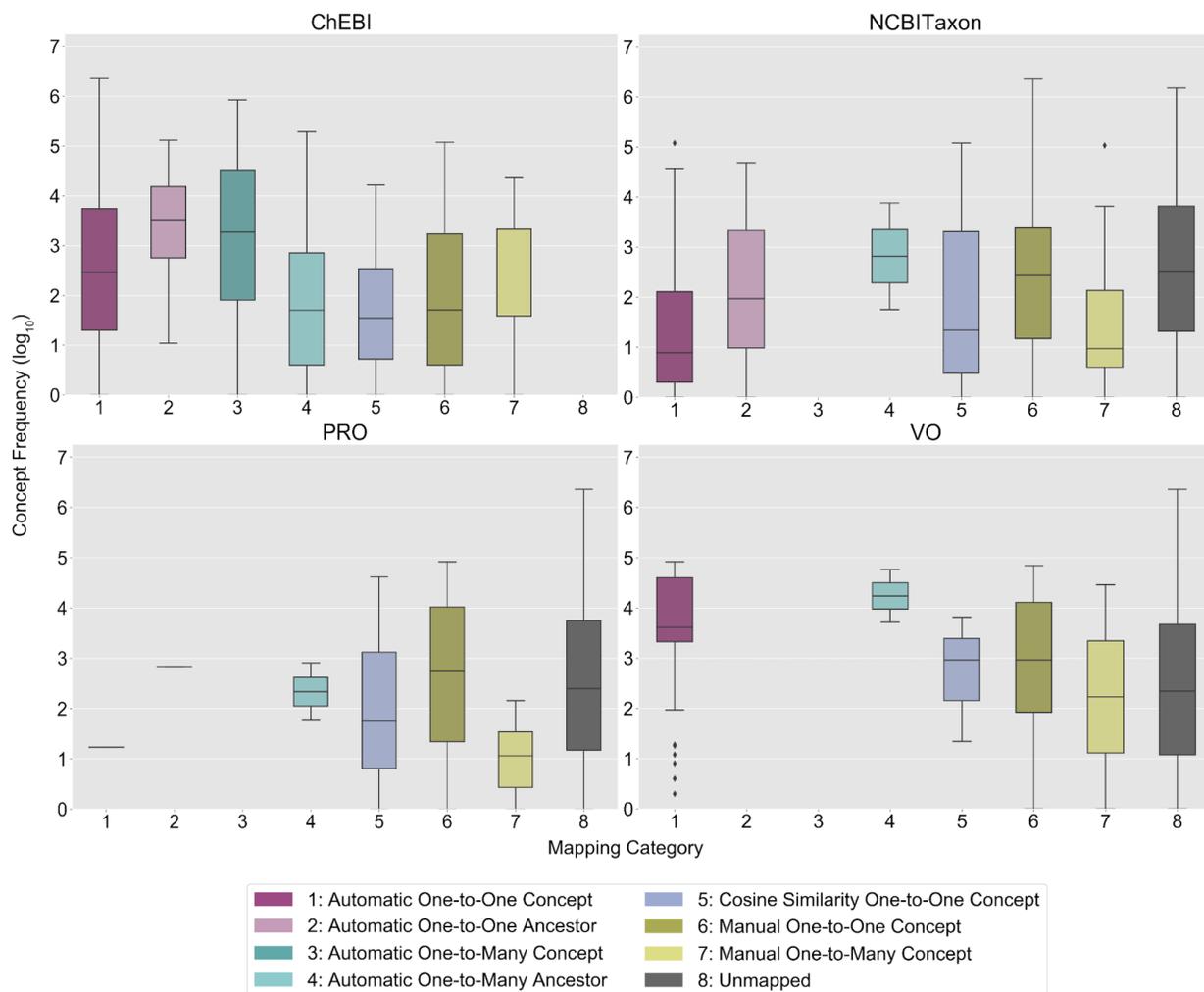

**Mapping Categories.** The frequency distributions of the *Concepts Used in Practice* by mapping category and OBO Foundry ontology are visualized in Figure 6. The majority of the automated mappings were one-to-one at the concept-level for *Concepts Used in Practice* (HPO: n=17, Uberon: n=1793, NCBITaxon: n=444, PRO: n=44, ChEBI: n=264, CL: n=182) and *Concepts Not Used in Practice* (HPO: n=3, Uberon: n=3589, NCBITaxon: n=444, PRO: n=12, ChEBI: n=515, CL: n=186). The majority of the manual mappings were one-to-one (HPO: n=3902, Uberon: n=406, NCBITaxon: n=2300, PRO: n=1267, ChEBI: n=1377, CL: n=319). Cosine similarity-scored concept embeddings enabled 113 HPO (*Concepts Used in Practice*: median 0.4, range 0.3-0.8; *Concepts Not Used in Practice:* median 0.4, range 0.3-0.9), 142 Uberon (*Concepts Used in Practice*: median 0.3, range 0.3-0.8; *Concepts Not Used in Practice:* median 0.4, range 0.3-0.7), 150 NCBITaxon (*Concepts Used in Practice*: median 0.4, range 0.3-0.7; *Concepts Not Used in Practice:* median 0.4, range 0.3-0.7), 132 PRO (*Concepts Used in Practice*: median 0.4, range 0.3-0.7; *Concepts Not Used in Practice:* median 0.4, range 0.3-0.6), 476 ChEBI (*Concepts Used in Practice*: median 0.4, range 0.3-1; *Concepts Not Used in Practice:* median 0.3, range 0.3-0.6), and 102 CL (*Concepts Used in Practice*: median 0.4, range 0.3-1; *Concepts Not Used in Practice:* median 0.4, range 0.3-1) mappings (Supplementary Figure 2c). On average, more evidence was found for mappings to *Concepts Used in Practice* than *Concepts Not Used in Practice* for HPO, Uberon, and PRO (HPO: 1.03 vs 1.02; Uberon: 2.3 vs 1.9; PRO: 1.1 vs 1; NCBITaxon: 1.3 vs 1.4; ChEBI: 2.7 vs 2.9; CL: 2.5 vs 2.8).

**Figure 6: Measurement Concept Frequency of Use in Clinical Practice by Mapping Category and OBO Foundry Ontology.**

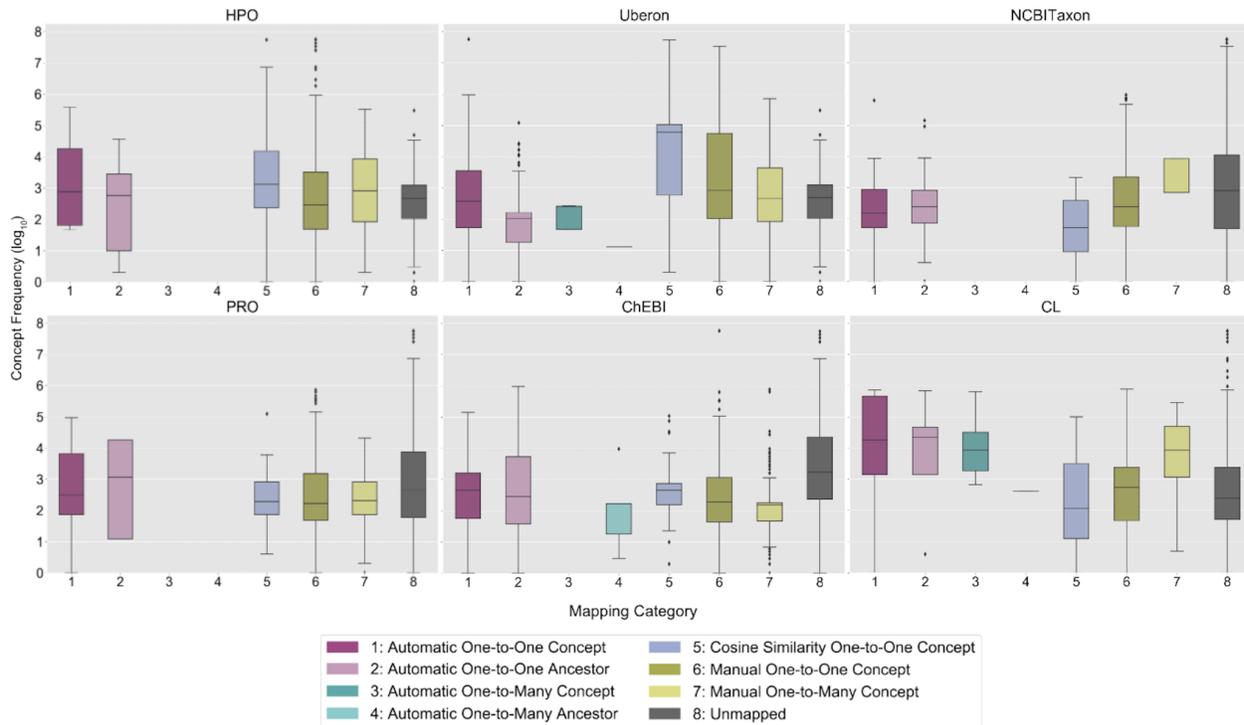

## 2.3 Validation

### 2.3.1 Accuracy

The goal of this task was to verify the accuracy of randomly selected sets of manual one-to-one and one-to-many OMOP2OBO mappings from each OMOP domain through domain expert review. Of the 2000 condition mappings, 73.9% were correct (n=1477). Of the 116 reviewed drug ingredient mappings, 70.7% (n=82) were correct. Upon review, it was found that 165 (31.6%) of the incorrect condition and 14 (41.2%) of the incorrect drug ingredient mappings could be improved by creating more specific mappings through adding new concepts to the OBO Foundry ontologies or by replacing multiple mappings to broad ancestor concepts with a single best representative ancestor concept. Measurement concepts were reviewed at the result-level using a survey and manual domain expert review. On the survey, 92.9% (n=251) of the mappings were found to be correct. Of the 1350 measurement results, 97.3% (n=1314) were correct.

In addition to expert review, each mapping was inspected at least twice by a member of the research team (TJC). If we assume that the automatic one-to-one mappings created using resources provided by the UMLS, OMOP CDM, and OBO Foundry ontologies are correct and exclude mappings that occur at the ancestor level (assuming those are too broad) and unmapped concepts, then the following concepts received at least one form of review: (I) Conditions: 18.4% of Mondo and 9.9% of HPO; (II) Drug Ingredients: 95.3% of NCBITaxon, 90.3% of VO, 85.3% of ChEBI, and 33.3% of PRO; and (III) Measurements: 79.2% of HPO, 50.8% of Uberon, 48.5% of CL, 12.7% of CHEBI, 10.6% of NCBITaxon, and 3.9% of PRO.

### 2.3.2 Generalization

The goal of this evaluation was to characterize the generalizability or coverage of concepts in the OMOP2OBO mapping set to a set of OMOP standard concepts that are commonly utilized in clinical practice. The Observational Health Data Sciences and Informatics (OHDSI) Concept Prevalence Study contains OMOP standard concepts that are commonly utilized in practice from several independent study sites across the OHDSI network (see Supplementary Table 3 for more information).[76–79] For this evaluation, we leveraged data (referred throughout the remainder of the manuscript as the "OHDSI Concept Prevalence Data") from 24 independent study sites, which included hospitals, academic medical centers, and claims databases. For this analysis, the OMOP2OBO mappings were filtered to identify all concepts with at least one valid mapping (i.e., excluding unmapped and not yet mapped concepts) across all of the ontologies mapped within each OMOP domain.

#### 2.3.2.1 Conditions

The OHDSI Concept Prevalence Data contained 62,335 condition concepts from 24 independent sites. The filtered OMOP2OBO mapping set contained 92,367 eligible concepts, which covered 92.5% (99.5% weighted coverage) of the OHDSI Concept Prevalence Data concepts (n=57,663 concepts; median 689, range 100-874,824,195). Of the remaining concepts, 34,704 were only found in OMOP2OBO (median 100, range 100-39,975) and 4672 were only found in the OHDSI Concept Prevalence Data (median 100, range 100-52,739,431). These findings are visualized in Figure 7a. OMOP2OBO concept coverage ranged from 93-99.7% across the 24 OHDSI Concept Prevalence Data sites. Supplementary Figure 3a presents the counts of OMOP concepts in the OHDSI Concept Prevalence Data by site. A Chi-Squared test of independence with Yate's correction revealed a significant association between the OHDSI Concept Prevalence Data sites and OMOP2OBO coverage ($\chi^2(23)$ =

**Figure 7: OMOP2OBO - Concept Prevalence Coverage.**

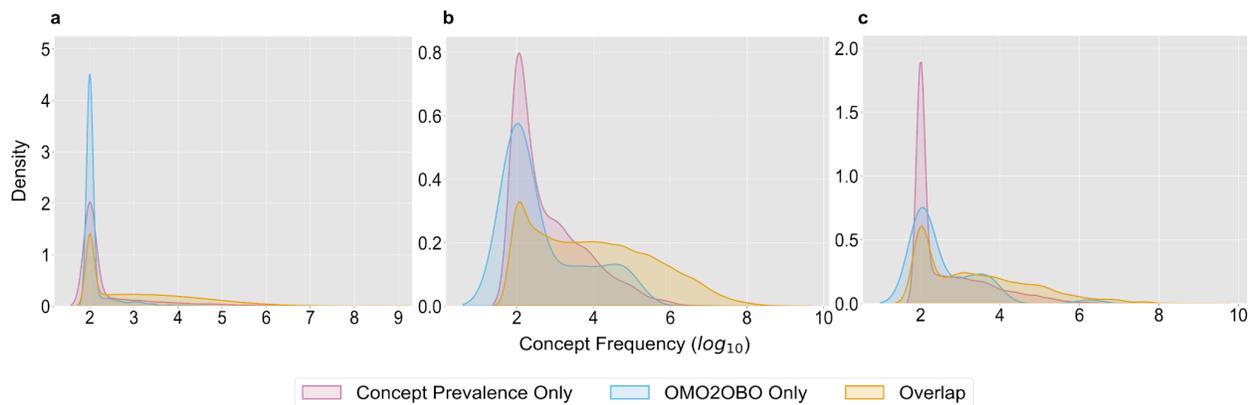

7559.1, p<0.0001). Post-hoc tests using Bonferroni adjustment confirmed that 32% of the pairwise OHDSI Concept Prevalence Data site comparisons had significantly different OMOP2OBO coverage (p<0.001 for all significant comparisons). The results of this analysis are visualized as a heatmap in Supplementary Figure 3b. The OMOP2OBO concept count by OBO Foundry ontology, data wave, and coverage type are shown in Supplementary Figure 4.

**Error Analysis.** Results for the 4672 (7.5%) OHDSI Concept Prevalence Data concepts missing from OMOP2OBO are visualized in Figure 8a. Roughly 7.9% (n=367) of concepts were accounted for using a newer version of the OMOP CDM and occurred in an average of 2.6 sites with a mean frequency of 27,412.3 (range 100-3,539,698.5). 90.6% (n=4231) of concepts purposefully excluded from the OMOP2OBO mapping set (i.e., no clear pathological or biological origin, not yet mapped, or were unable to be mapped) occurred in an average of 1.7 sites with a mean frequency of 6139.3 (range 100-8,254,186.5). The remaining concepts (1.6%; n=74) were truly missing and occurred in an average of 2.7 sites with a mean frequency of 5320.1 (range 100-100,483). The frequency of distributions of the covered OMOP2OBO concepts and Concept Prevalence concepts missing from OMOP2OBO in the OHDSI Concept Prevalence Data by site are visualized in Supplementary Figure 3c-d. The five most frequently occurring missing concepts are shown in Table 2. Domain expert review determined these concepts were likely missing due to differences in patient populations and coding practices. The domain experts identified comparable condition concepts in the OMOP2OBO mapping set.

**Figure 8: OMOP2OBO - Concept Prevalence Coverage Error Analysis.**

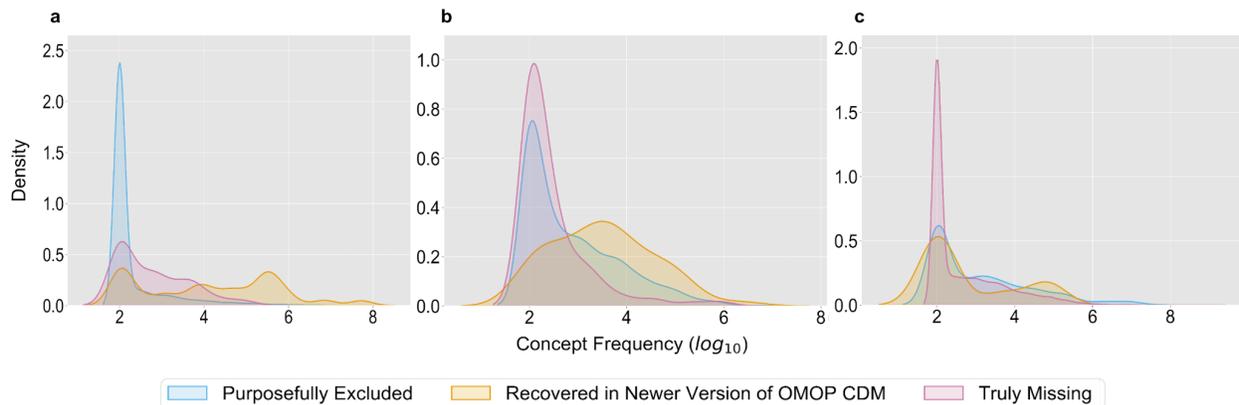

*2.3.2.2 Drug Ingredients*

The OHDSI Concept Prevalence Data contained 4588 drug ingredient concepts from 18 independent sites. The OMOP2OBO mapping set contained 8611 eligible concepts, which covered 87.9% (99.9% weighted coverage) of the OHDSI Concept Prevalence Data concepts (n=4037 concepts; median 7299, range 100-1,308,580,305). Of the remaining concepts, 4574 were only found in OMOP2OBO (median 100, range 100-69,311) and 551 were only found in the OHDSI Concept Prevalence Data (median 300, range 100-10,748,492). These findings are visualized in Figure 7b. OMOP2OBO concept coverage ranged from 91.2-98.4% across the 18 Concept Prevalence Study sites. Supplementary Figure 5a presents the counts of OMOP concepts in the OHDSI Concept Prevalence Data by site. A Chi-Squared test of independence with Yate's correction revealed a significant association between the OHDSI Concept Prevalence Data sites and OMOP2OBO coverage ($\chi^2(17) = 195.6$, $p<0.0001$). Post-hoc tests using Bonferroni adjustment confirmed that 22% of the pairwise OHDSI Concept Prevalence Data site comparisons had significantly different OMOP2OBO coverage ($p<0.001$ for all significant comparisons). The results of this analysis are visualized as a heatmap in

Supplementary Figure 5b. The OMOP2OBO concept count by OBO Foundry ontology, data wave, and coverage type are shown in Supplementary Figure 6.

**Error Analysis.** Results for the 551 (12%) OHDSI Concept Prevalence Data concepts missing from OMOP2OBO are visualized in Figure 8b. Roughly 0.9% (n=5) of concepts were accounted for using a newer version of the OMOP CDM and occurred in an average of 8.4 sites with a mean frequency of 51,732 (range 100-221,229.7). 82.8% (n=456) of concepts purposefully excluded from the OMOP2OBO mapping set (i.e., not yet mapped) occurred in an average of 3.9 sites with a mean frequency of 18,847.3 (range 100-1,077,258.9). The remaining concepts (16.3%; n=90) were truly missing and occurred in an average of 2.7 sites with a mean frequency of 3361.2 (range 100-175,551.3). The frequency of distributions of the covered OMOP2OBO concepts and Concept Prevalence concepts missing from OMOP2OBO in the OHDSI Concept Prevalence Data by site are visualized in Supplementary Figure 5c-d. The five most frequently occurring missing concepts are shown in Table 2. Domain expert review of these concepts found that they were likely missing as a result of hospital vendor differences or because they were a new high-risk biologic whose safety and efficacy had not yet been tested or confirmed for use in pediatric populations. The domain experts identified comparable drug ingredient concepts in the OMOP2OBO mapping set.

*2.3.2.3 Measurements*

The OHDSI Concept Prevalence Data contained 25,513 measurement concepts from 18 independent sites. The resulting OMOP2OBO mapping set contained 3828 eligible concepts (10,676 results), which covered 11.1% (67.7% weighted coverage) of the OHDSI Concept Prevalence Data concepts (n=2260 concepts; median 1355, range 100-1,465,815,430). Of the remaining concepts, 1208 were only found in OMOP2OBO (median 100, range 100-1,842,485) and 20,893 were only found in the OHDSI Concept Prevalence Data (median 109, range 100-1,219,846,862). These findings are visualized in Figure 7c. OMOP2OBO concept coverage ranged from 4.2-75% across the 18 OHDSI Concept Prevalence Data sites. Supplementary Figure 7a presents the counts of OMOP concepts in the OHDSI Concept Prevalence Data by site. A Chi-Squared test of independence with Yate's correction revealed a significant association between the OHDSI Concept Prevalence Data sites and OMOP2OBO coverage ($\chi^2$(17) = 3872.3, p<0.0001). Post-hoc tests using Bonferroni adjustment confirmed that 56% of the pairwise OHDSI Concept Prevalence Data site comparisons had significantly different OMOP2OBO coverage (p<0.001 for all significant comparisons). The results of this analysis are visualized as a heatmap in Supplementary Figure 7b. The OMOP2OBO concept count by OBO Foundry ontology, data wave, and coverage type are shown in Supplementary Figure 8.

**Error Analysis.** Results for the 20,893 (81.9%) OHDSI Concept Prevalence Data concepts missing from OMOP2OBO are visualized in Figure 8c. Roughly 0.1% (n=13) of concepts were accounted for using a newer version of the OMOP CDM and occurred in an average of 3.2 sites with a mean frequency of 9836.3 (range 100-29,098.2). 0.8% (n=158) of concepts purposefully excluded from the OMOP2OBO mapping set (i.e., not mapped test type, unspecified sample, or were unable to be mapped) occurred in an average of 5.2 sites with a mean frequency of 282,115.3 (range 100-14,317,951.9). The remaining concepts (99.2%; n=20,722) were truly missing and occurred in an average of 2.8 sites with a mean frequency of 218,874 (range 100-1,219,846,862). The frequency of distributions of the covered OMOP2OBO concepts and Concept Prevalence concepts missing from OMOP2OBO in the OHDSI Concept Prevalence Data by site are visualized in Supplementary Figure 7c-d. The five most frequently occurring missing concepts (reported as the average frequency across the 18 sites and number of sites with that concept) are shown in Table 2. Domain expert review of these concepts confirmed that they were likely missing due to inconsistencies in hospital use of LOINC, a finding that's been observed in literature.[80] The domain experts identified comparable measurement concepts in the OMOP2OBO mapping set.

*2.3.3 Clinical Utility*

Many patients with a genetic disease never receive a specific diagnosis, even after genetic sequencing.[81–84] Longitudinal EHR data has been used to identify patients with genetic disorders.[85–88] Inspired by the fact that most genetic diseases manifest as a recurring pattern of multiple symptoms or phenotypes affecting multiple organ systems,[86] the phenotype risk score (PheRS), which measures the similarity between an individual's diagnosis codes and phenotypic features of known genetic disorders, was developed.[85] While the PheRS has shown great promise for identifying patients with undiagnosed Mendelian disease from EHR data,[61] it requires mappings that link ICD codes to HPO concepts, which most EHRs do not contain. The existing mappings[61] developed to support PheRS were manually constructed, which may limit scalability when applied to new data.

The goal of this evaluation was to determine if the OMOP2OBO mappings could be used to facilitate the application of the PheRS to EHR data and to compare their performance to an existing set of validated manual mappings. For this analysis, the OMOP2OBO HPO mappings were compared to the ICD-HPO mappings[61] using data from the All of Us Research Program (AoU)[89]. The AoU Data were selected for this task because it provides access to a large sample of EHR data and genetic testing results (see Supplementary Table 3 for additional details on this data source). Five genetic diseases (and their associated genes) for which diagnosis codes have been found to be of high positive predictive value in EHRs,[61] were examined: Marfan syndrome (*FBN1* and *TGFBR1*), multiple endocrine neoplasia (*MEN1* and *RET*), neurofibromatosis (*NF2*), paragangliomas (*SDHAF2*, *SDHB*, *SDHC*), and tuberous sclerosis (*TSC1*, *TSC2*). These diseases were associated with 2257 unique phenotypic features (HPO codes). When querying AoU data to identify patients who had at least one of these phenotypic features, the ICD-HPO mappings (n=7815 ICD codes) took ~30 minutes to complete and returned 210,718 patients and the OMOP2OBO mappings (n=3783 OMOP concepts) took ~10 minutes to complete and returned 209,342 patients. Of the 208,831 patients found in common, 1887 were only identified by the ICD-HPO mappings, and 601 patients were only identified by the OMOP2OBO mappings. When the PheRS was applied to patients from both mappings they were found to be highly correlated ($r^2>0.6$ across all diseases). This suggests that the patients returned by both mappings are similar.

For additional validation, case-control studies using only the OMOP2OBO mappings were performed: Marfan syndrome (131 cases and 63,086 controls), multiple endocrine neoplasia (86 cases and 72,150 controls), neurofibromatosis (255 cases and 65,256 controls), paraganglioma (105 cases and 65,256 controls), and tuberous sclerosis (38 cases and 58,555 controls). The results of these studies are shown in Supplementary Table 8 and the distributions of PheRS scores for cases and controls for each of the five diseases are visualized in Supplementary Figure 9. As shown in this figure, PheRS were higher for cases than controls across all examined diseases. These results are further supported by one-sided Wilcoxon rank sum tests, which indicated that the PheRS were significantly higher for cases than controls ($p<0.001$ for all diseases). Collectively, these results support the use of OMOP2OBO mappings as a scalable alternative to an existing set of validated manual mappings for use with PheRS to aid in the systematic identification of patients who might benefit from genetic testing.

## 3.0 DISCUSSION

In this paper we present OMOP2OBO, an algorithm that semantically aligns conditions, drug ingredients, and measurement results from standard vocabularies in the OMOP CDM to OBO Foundry ontologies. Using OMOP2OBO, we built mappings for 92,367 condition, 8615 drug ingredient, and 10,673 measurement result concepts to ontology concepts representing 9636 diseases, 6309 phenotypes, 83 anatomical entities, 2704 organisms, 4261 chemicals, 132 vaccines, and 272 proteins. The mappings were evaluated on accuracy, generalizability, and clinical utility. For the first task, a panel of 10 domain experts reviewed subsets of the manually-derived mappings from each of the OMOP domains and found that 73.9% of the condition, 70.7% of the drug ingredient, and 92.9% of the measurement result mappings were correct. For the second task, we examined the generalizability of the concepts and found that 99.5% of conditions, 99.9% of drug ingredients, and 68% of measurement results overlapped with concepts used in clinical practice from 24 independent hospitals and claims databases. For the final task, we compared OMOP2OBO HPO mappings to an existing set of validated manual mappings when used to identify patients with five rare genetic diseases using data from the AoU Research Program. Queries using the OMOP2OBO mappings identified 99.3% of the patients returned by the validated manual mappings using fewer codes and one-third of the time.

To the best of our knowledge, the OMOP2OBO mappings are the largest set of publicly available mappings between clinical vocabularies and OBO Foundry ontologies. The OMOP2OBO algorithm can easily be incorporated into existing clinical workflows and presents new opportunities to advance EHR-based deep phenotyping (recently published examples are described below in the **Applications** section).

## 3.1 Related Work

Existing work to develop mapping sets and mapping algorithms has largely focused on using ontologies to improve the phenotyping of specific diseases (e.g., infectious disease,[90] rare diseases,[91,92] and cancer[93]) and for the investigation of specific biological (e.g., glycobiology[94]) and clinical domains (e.g., laboratory test results[62] and medical diagnoses[64,95]). Our work is most similar to LOINC2HPO,[62] which we have expanded in our current mapping set (with annotations to five additional OBO Foundry ontologies). OMOP2OBO complements existing phenotyping efforts like the Electronic Medical Records and Genomics or eMERGE Network[96] and the AoU Research Program,[89] by providing access to resources not currently available in EHRs and opportunities to improve the semantic interoperability of definitions through alignment to the OBO Foundry ontologies.

A portion of the mappings that are automatically derived by OMOP2OBO overlap with existing mappings provided by the OMOP CDM, the UMLS, and the OBO Foundry ontologies. The UMLS and OMOP CDM each align more than 200 vocabularies. At the time of our analysis (and determined using the same data), only the UMLS provided mappings to an OBO Foundry ontology, which included: the Gene Ontology (67,807 CUIs covering 69 vocabularies and an average of 166.9 codes), HPO (16,154 CUIs covering 91 vocabularies and an average of 1668.7 codes), and NCBITaxon (1,776,212 CUIs covering 55 vocabularies and an average of 3236.1 codes). Of these mappings, only the HPO and NCBITaxon are relevant to our work. Of the 1,776,212 CUIs aligned to NCBITaxon, 1128 were mapped to LOINC and 138 were mapped to RxNorm covering 0% of the measurement and 1.1% of the drug ingredient concepts in the CHCO OMOP Database, respectively. Of the 16,154 CUIs aligned to HPO, 993 were mapped to LOINC and 18,212 were mapped to SNOMED-CT covering 0% of the measurement and 4.2% of the condition concepts in the CHCO OMOP Database, respectively. Similar to the OMOP CDM and the UMLS, some of the OBO Foundry ontologies provide mappings to vocabularies in these resources. Collectively, the eight OBO Foundry ontologies used in this work provided 489,794 unique database cross-references from 179 unique data sources. Of these, only the HPO (11,616 ontology concepts to 19,569 codes from 16 data sources), Mondo (22,110 ontology concepts to 159,918 codes from 45 data sources), CL (949 ontology concepts to 1376 codes from 29 data sources), and Uberon (10,865 ontology concepts to 51,322 codes from 91 data sources) mappings were relevant to our work. Of the 19,569 HPO and 159,918 Mondo database cross-references only 3.6% and 15.6% mapped to a condition concept in the CHCO OMOP Database, respectively.

These findings highlight that while there are some existing mappings between the resources that OMOP2OBO aligns, at best, they covered only ~15% of the OMOP concepts that we aimed to map supporting the need for its development. Further, it should be noted that the vast majority of the mappings provided by the OMOP CDM, UMLS, and OBO Foundry ontologies are simple one-to-one mappings. While OMOP2OBO also contributes one-to-one mappings, it provides more complex one-to-many mappings.

## 3.2 Applications

The OMOP2OBO mappings have been used to characterize differences in definitions of long COVID,[97] generate long COVID phenotypes,[98,99] and improve the categorization and prediction of psychiatric diseases among patients with long COVID.[100] Additionally, our recent work in pediatric rare disease subphenotyping demonstrated that patient representations constructed from the OMOP2OBO mappings produced more clinically meaningful clusters than representations built using OMOP concepts alone.[101] We further demonstrated the value of the mappings by leveraging them to successfully integrate external gene expression data from an independent sample of pediatric patients resulting in more clinically-meaningful and biologically-actionable phenotypes than those generated using only clinical data.

One potential use of OMOP2OBO is to aid in the alignment of patient data to ontologies in the Global Alliance for Genomics and Health's Phenopacket schema,[102] which was designed to support the global exchange of computable patient-level phenotypic information. This work was discussed during the 2021 ELIXIR European BioHackathon.[103]

## 3.3 Limitations and Future Work

### 3.3.1 Limitations

OMOP2OBO has not been optimized for performance; all possible ancestors are mapped when unable to generate a mapping at the concept-level. A prioritization strategy would significantly improve performance. OMOP2OBO does not take advantage of all of the knowledge available in the UMLS. Leveraging information in the mapping and hierarchy tables could improve the automatically mapped concepts and would enable use of other UMLS-aligned resources like the SemMedDB.[104] We only evaluated the accuracy of a small subset of the manual mappings. It is important to evaluate the remaining manually derived mappings as well as to provide citations from the resources from which they were derived. The *Accuracy* evaluation revealed limitations of our expert review procedures; some of the experts experienced challenges when trying to use the OBO ontologies, which may have negatively impacted the results. Providing better training and offering outcomes other than correct/incorrect should be considered. Finally, OMOP standard clinical vocabularies are also dependent upon a large set of CDM-specific mappings and may be subject to similar errors as our mappings.

### *3.3.2 Future Work*

There are two primary challenges that remain given the initial development of the OMOP2OBO algorithm and mapping set. The first challenge is to establish procedures and build infrastructure to enable community sharing, monitoring, and updating of the mappings. While the GitHub repository for the OMOP2OBO currently contains policies for contributing to the mapping algorithm, we have yet to establish an infrastructure or policies for the mappings. Future opportunities include the adoption of a system like the one utilized by the Bioregistry (https://bioregistry.io/).[105] The Bioregistry provides extensive governance policies and templates, which make it easy to incorporate new and modify existing identifiers. They also developed a robust, semi-automated infrastructure that facilitates review by the maintainers and triggers rebuilds of the registry anytime changes are made. To improve the shareability of the mappings, we'd also like to extend the mapping output formats to include Semantic Web standards like RDF/XML and the Simple Standard for Sharing Ontological Mappings or SSSOM.[106] In addition to creating a system like the Bioregistry, future work may include adoption and adaptation of OBO Foundry protocols for ontology development and maintenance.[60,107]

The second challenge is to improve and expand the evaluation of the algorithm and the mapping set. The UMLS, OMOP CDM, and the OBO Foundry ontologies provide mappings between clinical vocabularies and ontologies, which are automatically- or manually-derived (e.g., mappings between source and standard vocabulary concepts, mappings between clinical vocabularies and ontologies, and/or database cross-references mapped to ontology concepts).

While the OMOP2OBO algorithm leverages these mappings (i.e., leveraging source codes mapped to standard concepts), verifying the quality of existing mappings was not within the scope of the current work. Currently, no modules in the OMOP2OBO algorithm verify the quality of existing mappings used by OMOP2OBO or mappings generated by it. This should include resources to validate automatic mappings as their accuracy depends upon the quality of the resources from which they were built, and ontologies are subject to a variety of errors.[108–110] To do this, we might leverage pretrained language models and/or develop new machine learning models using trusted resources (e.g., the scientific literature) to verify the database cross-references provided by the OBO Foundry ontologies, UMLS, and OMOP CDM database prior to running OMOP2OBO.

## 4.0 METHODS

OMOP2OBO is open source (https://github.com/callahantiff/OMOP2OBO), available on PyPI (https://pypi.org/project/omop2obo/), and includes an interactive dashboard that summarizes the current mapping set (http://tiffanycallahan.com/OMOP2OBO_Dashboard/). We also created a dedicated Zenodo Community, which provides access to data, mappings, and presentations (https://zenodo.org/communities/omop2obo). A list of the acronyms used in this paper are provided in Supplementary Table 1 and the resources used by the OMOP2OBO algorithm and mappings are described in Supplementary Table 2.

### 4.1 OMOP2OBO Algorithm

*4.1.1 Algorithm Resources*

Although it is possible to apply the OMOP2OBO algorithm to any clinical vocabulary, the OMOP CDM was selected because of its rich data representation, standard vocabularies (and hierarchies) and the mappings it provides to more than 200 commonly used clinical vocabularies. To increase the coverage of the resources and the potential of an automatic mapping, OMOP2OBO leverages the National Library of Medicine's UMLS (MRCONSO and MRSTY tables [2020AA version[111]])[74]. These data are used to annotate each OMOP concept with a UMLS CUI and a Semantic Type.[75] Additionally, the mappings provided by the MRCONSO table are used to enhance existing database cross-reference mappings provided by OMOP and the OBO Foundry ontologies (both described in detail in the **Input Data Used to Create OMOP2OBO Mappings** section).

*4.1.2 Algorithm Overview*

The OMOP2OBO algorithm (Figure 1) consists of the following three components: (I) Process Input Data; (II) Map OMOP Standard Vocabulary Concepts to OBO Foundry Ontology Concepts; and (III) Synthesize and Process Mapping Results and Output Mappings. Each component is described in detail below.

    **Component 1: Process Input Data.** The algorithm takes as input a table of OMOP concepts and a list of one or more OBO Foundry ontologies. For both types of data, the algorithm expects concept or class identifiers, source codes or database cross-references, labels, synonyms, and ancestor concepts or classes. While the algorithm expects a table of input OMOP concepts (due to the private nature of clinical data, the algorithm does not assume a direct database connection is possible), it automatically downloads the OBO Foundry ontologies using OWLTools (April 06, 2020 release).[112]

    **Component 2: Map OMOP Vocabulary Concepts to OBO Foundry Ontology Concepts.** This component is designed to automatically map or align OMOP concepts to OBO Foundry ontology concepts. The algorithm includes several different approaches, prioritizing those that result in high confidence mappings. This component includes concept alignment and concept embedding.

- <u>Concept Alignment</u>. Exact-string matches between OMOP and OBO Foundry ontology concept labels, definitions, and synonyms are obtained. Prior to alignment, the label and synonym fields are both made lowercase. This step also obtains exact matches between OMOP standard concepts and source codes to OBO Foundry ontology database cross-references. To increase the likelihood of finding a match, the OMOP standard concepts and source codes are first merged with terminologies in the UMLS using core functionality from OHDSI Ananke,[113] a program developed to align OMOP concepts to UMLS CUIs. Prior to performing this alignment, the OMOP standard concepts and source codes and the OBO Foundry ontology database cross-references are normalized using a custom dictionary (source_code_vocab_map.csv[114]). This resource ensures that concepts referenced by the same code using different prefixes or symbols can be aligned (e.g., SNOMED-CT:1234567 and sctid:1234567).

- <u>Concept Embedding</u>. Using scikit-learn,[115] a bag-of-words (BoW)[116] vector space model with term-frequency inverse-document frequency (TF-IDF)[117] and L2 normalization is used to learn concept embeddings for all OMOP and OBO Foundry ontology concepts and concept ancestors label and synonym text strings. While the BoW model was used because it is easy to understand and has shown great success when applied to EHR data and when used to align biomedical ontologies,[118,119] any language or embedding model could be utilized. The BoW model is implemented as an NxM document-term matrix with one row per input string and one column for each tokenized word appearing in the universe of all input strings. The value of each cell in the matrix is the normalized frequency each word occurred in each input string (using TF-IDF normalization). Prior to building the model, all text fields are made lowercase, stop words are removed using the wordnet list from Python's NLTK library,[120] white spaces are removed, and word-level tokenization and lemmatization are applied. After learning the model, a final embedding is constructed for each input string by aggregating the constituent concept embeddings. Cosine similarity is used to compute scores between all pairwise combinations of OMOP and OBO Foundry concept embeddings. Given that each OMOP and OBO concept can have a label and one or more synonyms, only the single highest-scoring pairwise comparison is selected for the final mapping. Cosine similarity scores range from 0-1, where a score of one indicates a greater match between the embedding pairs. To improve the efficiency of this process, only the top 75% of pairs with scores >=0.25 are output, which was decided after visualizing the score distribution using a histogram. All thresholds and cut-offs are customizable. Concept embeddings are created for all OMOP concepts, regardless of whether or not they were automatically mapped by a prior Component. All remaining unmapped concepts require manual curation.

**Component 3: Synthesize and Output Mapping Results.** The mapping results from the prior component are post-processed to include a mapping category and human-readable evidence. The mapping category is constructed by combining the following elements: (I) one or more OBO Foundry ontology identifiers and labels; (II) mapping logic applied to specify semantics when there are multiple ontology concepts (i.e., "and", "or") or to denote negation (i.e., "not"); (III) a mapping category derived from the mapping approach (e.g., automatically determined using an algorithm or manually derived by a human annotator), cardinality (i.e., one-to-one aligning a single OMOP concept to a single OBO Foundry ontology concept or one-to-many aligning a single OMOP concept to one or more OBO Foundry ontology concepts), and level (i.e., mapping to the OMOP concept directly or to one of its ancestors); and (IV) mapping evidence represented as a pipe-delimited string that denotes all resources that support the mapping (i.e., the exact string matches between labels and synonyms, source codes and database cross-reference alignments, and other sources supporting a mapping like scored heuristics and references from manual review). Supplementary Table 4 provides additional details on and examples of the mapping categories. Post-processed mappings are serialized and able to be output to a variety of file types, like flat file, database dump, or RDF/XML file.

### 4.2 Input Data Used to Create OMOP2OBO Mappings

The OMOP2OBO mappings were constructed from two data sources: (i) the CHCO OMOP

Database and (II) OBO Foundry ontologies (both are described in detail below). Figure 2 includes example mappings and illustrates how the OBO Foundry ontologies were used to map OMOP concepts from each domain. Supplementary Table 4 provides additional details on and examples of the mapping categories. Supplementary Table 3 provides descriptions of the clinical data sources used to build and validate the OMOP2OBO mappings.

*4.2.1 OMOP Data*

The OMOP2OBO mappings were constructed using data from the CHCO OMOP Database, a de-identified database that contained data from more than 6 million pediatric patients. The CHCO OMOP Database is stored within University of Colorado Anschutz Medical Campus Health Data Compass' Health Insurance Portability and Accountability Act compliant Google Cloud-based infrastructure (created in October 2018).[121] The data conformed to the structure defined by the National Pediatric Learning Health System (PEDSnet) OMOP CDM v3.0, which is an adaptation of the OMOP CDM version 5.0.[56,65] Use of these data was approved by the Colorado Multiple Institutional Review Board (#15-0445).

Concept lists were derived from standard OMOP vocabularies (i.e., SNOMED-CT[54] [v20180131], RxNorm[55] [v20180507], and LOINC[53] [v2.64]) from the Condition Occurrence, Drug Exposure (at the drug ingredient level), and Measurement tables, respectively. For each concept set, metadata were extracted from the OMOP CDM including concept codes (i.e., codes from each standard vocabulary), labels, synonyms, and ancestor concepts (codes, labels, and synonyms were also extracted for each concept ancestor). Concept lists were organized into two data waves according to whether or not they had been used in clinical practice (i.e., *Concepts Used in Practice* and *Concepts Not Used in Practice)*. As manual annotation requires significant resources, only concepts from the first data wave (i.e., *Concepts Used in Practice*) were manually mapped. Prior to constructing the concept lists, Condition and Measurement data were preprocessed to ensure the mapping process was robust and reproducible.

*4.2.1.1 Conditions*

For *Concepts used in Practice*, UMLS Semantic Types were used to identify all concepts that had a clear pathological or biological origin. All remaining concepts (e.g., accidents, injuries, external complications, and findings without clear interpretations) were marked as unmapped and the reason for exclusion was provided in the evidence field. The Semantic Types were also used to group OMOP concepts such that those typed as "Findings" or "Signs and Symptoms" were treated as phenotypes and only mapped to HPO and concepts typed as "Disease or Syndrome" were only mapped to Mondo. For *Concepts Not Used in Clinical Practice*, all possible automatic mappings were obtained and concepts which were unable to be mapped automatically were marked as unmapped and "NOT YET MAPPED" was provided as the mapping evidence. This same approach was applied to drug ingredients.

*4.2.1.2 Measurements*

For all measurement concepts, a scale and result type were created. The scale (i.e., ordinal, nominal, quantitative, qualitative, narrative, doc, and panel) of each measurement was identified from the OMOP CDM or by parsing the concept synonym field. For all *Concepts Used in Practice,* reference ranges were used to determine the result type; concepts with numeric reference ranges were typed as "Normal/Low/High" and concepts with reference ranges that included "positive" or "negative" were typed as "Positive/Negative". *Concepts Not Used in Practice* with an ordinal scale or with synonyms that contained the words "presence" or "screen" were typed as "Positive/Negative". Concepts with a quantitative scale were typed as "Normal/Low/High". All other scale types were typed as "Unknown Result Type". While it is possible to infer the result type from the scale type (e.g., all concepts with a quantitative scale have result type "Normal/Low/High" and all concepts with an ordinal scale have result type "Positive/Negative"), our approach aimed to maximize the inclusion of concepts from all scale types. Mappings were created for each result type using the procedures defined by LOINC2HPO[62]; results were annotated with respect to their result type:

- <u>Concepts with result type "Normal/Low/High"</u>. For example, *Corticotropin [Mass/volume] in Plasma --4th specimen post XXX challenge* (LOINC:12460-2). Results above the reference range are mapped to *Increased Circulating ACTH Level* (HP:0003154). Results below the reference range are mapped to *Decreased Circulating ACTH Level* (HP:0002920). Results within the reference are mapped to *Abnormality of Circulating Adrenocorticotropin Level* and logically negated (NOT HP:0011043).

- <u>Concepts with result type "Positive/Negative"</u>. For example, *Amphetamine [Presence] in Urine by Screen Method* (LOINC:19343-3). Positive results are mapped to *Positive Urine Amphetamine Test* (HP:0500112). Negative results are mapped to *Positive Urine Amphetamine Test* and logically negated (NOT HP:0500112).

Also consistent with the procedures adopted by LOINC2HPO, all concepts lacking sufficient detail (i.e., non-specific body substances) were marked as unmapped and "Unspecified Sample" was provided as the mapping evidence.

**LOINC2HPO Extensions.** The initial set of measurement concepts was supplemented with LOINC2HPO annotations,[62] which were downloaded on August 2, 2020 from the LOINC2HPO annotation Github repository.[122] OMOP2OBO expands the LOINC2HPO mappings by including the measurement substance (i.e., body fluids, tissues, and organs via Uberon), the entity being measured (i.e., chemicals, metabolites, or hormones via ChEBI; cell types via CL; and proteins via PRO), and the species of the measured entity (i.e., organism taxonomy via NCBITaxon). All modifications to the original LOINC2HPO annotations were recorded in the mapping evidence field, enabling users to easily identify when an original LOINC2HPO annotation had been updated. All LOINC concepts in the LOINC2HPO mappings that were not used at least once in clinical practice in the CHCO pediatric OMOP Database were categorized as a *Concept Not Used in Practice*.

*4.2.2 OBO Foundry Ontologies*

OBO Foundry ontologies were selected under the advice of several clinicians, molecular biologists, and professional OBO Foundry biocurators to cover the following domains: diseases (Mondo[67] [v2020-09-14]), phenotypes (HPO[63] [v2020-08-11]), anatomical entities (CL[69] [v2020-05-21], Uberon[68] [v2020-06-30]), organisms (NCBITaxon[70] [v2020-04-18]), chemicals (ChEBI[71] [v191]), vaccines (VO[72] [v1.1.102]), and proteins (PRO[73] [v61.0]). Similar to the clinical concepts, each ontology was queried to obtain labels, definitions, synonyms (including synonym type), and database cross-references. All OBO Foundry ontologies were downloaded in September 2020 using OWLTools (April 06, 2020 release).[112]

## 4.3 Mapping Evaluation

The OMOP2OBO mappings were evaluated by assessing their accuracy, generalizability, and clinical utility.

*4.3.1 Accuracy*

Automatic mappings are created from exact alignments between resources available in the OMOP CDM and the OBO Foundry ontologies and thus are assumed to be accurate and high-confidence mappings. The goal of this evaluation was to evaluate the accuracy of a portion of the manually-derived mappings. For conditions and drug ingredients, of all manual mappings (including one-to-one and one-to-many), 20% were randomly selected for manual review (n=2,000 conditions; n=116 drug ingredients) by a practicing resident physician and clinical pharmacist, respectively.

    Measurement mappings are significantly more complex as they require interpreting lab test results and annotating the source of the sample (e.g., bodily fluid, anatomical entity, or cell type), entity being measured (e.g., chemical or cell type), and organism of the measured entity. While annotating the samples and entities is straightforward, interpreting lab tests results and aligning them to HPO concepts can be challenging. As a result, only the HPO mappings were

evaluated by domain experts. These mappings were evaluated in two ways: (i) <u>Survey</u>. A subset of the mappings (n=270) were independently validated by five domain experts including three practicing pediatric clinicians, a PhD-level molecular biologist, and a master's-level epidemiologist using a Qualtrics Survey.[123] Any mapping that did not meet agreement by at least one clinician and both the biologist and the epidemiologist were re-evaluated by the most senior clinician. These mappings were also vetted on the LOINC2HPO GitHub tracker[124] by members of the biocuration team. (ii) <u>Biocurator Validation</u>. A random subset of 1350 measurement results were manually verified by an OBO Foundry biocurator.

All of the manual mappings were derived by a member of the research team who at the time of the analysis was a Computational Biology PhD candidate with Masters-level training in epidemiology and biostatistics (TJC). As this individual does not have specialized medical or pharmacological training, it is assumed that these mappings may contain errors. Additional details are provided on GitHub (https://github.com/callahantiff/OMOP2OBO/wiki/Accuracy).

### *4.3.2 Generalizability*

The generalizability of the OMOP2OBO mappings were examined using the OHDSI Concept Prevalence Study data.[76–79] The Concept Prevalence study provides data on the frequency of OMOP concept usage in clinical practice across several independent sites in the OHDSI network. In addition to the Concept Prevalence Study sites, data were obtained from two independent academic medical centers, bringing the total number of sites to 24. None of the 24 sites overlapped with the site that was used to generate the OMOP2OBO mappings. Consistent with the Concept Prevalence Study procedures, all concepts from the OMOP CHCO Database occurring fewer than 100 times were assigned a count of 100. For all other analyses, the true range of counts in the OMOP CHCO Database were utilized. The OMOP2OBO mappings were filtered to remove all concepts without at least one ontology mapping. Coverage of all standard OMOP concepts in the OMOP2OBO mapping set was assessed by identifying: (I) concepts that existed in the OMOP2OBO set and in at least one Concept Prevalence Study site (i.e., Overlap); (II) concepts only present in the OMOP2OBO set (i.e., OMOP2OBO Only); and (III) concepts only present in the Concept Prevalence Study set (i.e., Concept Prevalence Only).

**Error Analysis.** An error analysis was performed to examine the Concept Prevalence Only concept set. Three scenarios were examined: (I) <u>Recovered in Newer Version of CDM</u>: concepts that could be recovered using a newer version of the OMOP CDM (v5.3.1; 02/25/2022); (II) <u>Purposefully Excluded</u>: concepts without clear pathological or biological origin that were purposefully excluded from the OMOP2OBO mapping set; and (III) <u>Truly Missing</u>: concepts that could not be accounted for using the prior two scenarios. For all scenarios, concept frequency within the Concept Prevalence Study sites was used as a measure of concept importance. Findings from each scenario were reviewed by a practicing resident physician and a clinical pharmacist. See GitHub for Additional details (https://github.com/callahantiff/OMOP2OBO/wiki/Generalizability).

### *4.3.3 Clinical Utility*

The clinical utility of the OMOP2OBO mappings was compared to an existing set of validated manual mappings (ICD-HPO mappings[61]) when used to identify undiagnosed rare disease patients. For this analysis, AoU Data[89] was selected because it provides access to a large sample of EHR data with genetic testing results. For this evaluation, the version 6 build was used, which contained data from ~630 sites on more than 528,000 patients.[89] Five genetic diseases for which diagnosis codes have been found to be of high positive predictive value in EHRs[61] were selected, which included: Marfan syndrome, multiple endocrine neoplasia, neurofibromatosis, paraganglioma, and tuberous sclerosis. These diseases are associated with 11 of the 73 American College of Medical Genetics and Genomics (ACMG) secondary finding genes (ACMG-73; v3.0), which have specific mutations known to cause disorders, have well-defined phenotypes, and are clinically actionable.[125] The diseases and associated genes included: *FBN1* and *TGFBR1* (Marfan syndrome); *MEN1* and *RET* (multiple endocrine neoplasia); *NF2* (neurofibromatosis); *SDHAF2*, *SDHB*, and *SDHC* (paragangliomas); and *TSC1*, *TSC2* (tuberous sclerosis). Using the Online Mendelian Inheritance in Man (OMIM)[126] database and the HPO gene annotation table,[127] each gene and its corresponding set of phenotypic features were aligned to the HPO. To calculate the phenotypic burden of each genetic disease, HPO mappings to OMOP condition concepts from OMOP2OBO (v2.0.0 beta) and ICD concepts from a validated set of ICD-HPO mappings[61] were queried against the AoU data. PheRS for each gene were then calculated for patients from each the OMOP2OBO and Phecode mapping sets. The PheRS[85] is an algorithm used to identify patients with phenotypic features that are clinically similar to OMIM[126] Mendelian profiles but who lack formal diagnosis and has demonstrated utility for identifying underdiagnosed rare disease patients using only EHR data.[61,85] For this evaluation, the standardized PheRS was used because it is easier to interpret and reduces noise when it is suspected that a large number of phenotypes will overlap between cases and controls.[85] The OMOP2OBO and ICD-HPO mappings were compared and evaluated on time to complete the query against the AoU Data and differences in the returned patient cohorts. As validation, case-control studies were performed for each of the five diseases using the patients returned from the OMOP2OBO mappings. Cases were defined as patients with at least two occurrences of a relevant diagnosis code and control patients had no instances of these codes. Cases and controls were matched on age, sex, and length of EHR record. For each disease, a one-sided Wilcoxon rank sum test was performed in order to determine if PheRS were significantly higher for cases than controls. Results were verified by a PhD-level Epidemiologist specializing in genetics (CZ).

### 4.4 Statistics and Technical Specifications

OMOP2OBO was developed using Python 3.6.2 on a single machine with 8 cores and 16GB of RAM. All code and project information are publicly available and detailed on GitHub

(https://github.com/callahantiff/OMOP2OBO). The OMOP2OBO (v1.0) mappings are publicly available from Zenodo.[128–130] The OMOP2OBO Mapping Dashboard was built with R (v4.2.1) using Rmarkdown (v2.14) and flexdashboard (v0.5.2).

Descriptive and inferential statistics were performed to evaluate the data available for mapping and the OMOP2OBO mapping set. Chi-squared tests of independence with Yate's correction were used to: (I) assess differences in the proportions of metadata available from each OBO Foundry ontology; and (II) assess differences in the proportions of mapped concepts between OHDSI Concept Prevalence sites. Post-hoc tests using Bonferroni adjustment to correct for multiple comparisons were performed for significant omnibus tests. Analyses were performed in Jupyter Notebooks (v6.1.6) using the scipy (v1.4.1), statsmodels (v0.12.1), statistics (v1.0.3.5), and numpy (v.1.18.1) libraries. Visualizations were created using matplotlib (v.3.3.2). The *Clinical Utility* evaluation was performed in the AoU Researcher Workbench[131] using R (v4.1.2) and Python (v3.7). Analyses were performed on a machine with 16 CPUs and 60GB of memory.

**DATA AVAILABILITY**

Supplementary Table 2 lists the resources used by the OMOP2OBO algorithm. The MRCONSO and MRSTY tables (2020AA) require a license and are available through the UMLS (https://www.nlm.nih.gov/research/umls/licensedcontent/umlsknowledgesources.html). The data used to build and validate the OMOP2OBO mappings (v1) are described in Supplemental Table 3. The OMOP concepts are available for download through Athena (https://athena.ohdsi.org/). The OBO Foundry ontologies are publicly available (https://obofoundry.org/). The OMOP2OBO (v1.0) mappings are publicly available and can be downloaded from Zenodo: Conditions (https://doi.org/10.5281/zenodo.6774363); Drugs (https://doi.org/10.5281/zenodo.6774401); and Measurements (https://doi.org/10.5281/zenodo.6774443).

**CODE AVAILABILITY**

OMOP2OBO is publicly available through GitHub (https://github.com/callahantiff/OMOP2OBO) and PyPI (https://pypi.org/project/omop2obo/).

**ACKNOWLEDGEMENTS**

This work was supported by funding from the National Library of Medicine (T15LM009451) to LEH and (T15LM007079) to GH and in part by the Intramural Research Program of the National Human Genome Research Institute (ZIA HG200417) to JCD. The authors thank colleagues at the Health Data Compass Warehouse and the OMOP2OBO and Machine Learning Working Groups at the National COVID Cohort Collaboration (NCATS U24TR002306) for piloting testing the mappings. The authors would also like to thank Drs. Paul Schofield (University of Oxford) and members of the Denny Lab (National Human Genome Research Institute) for reviewing the mappings.

**AUTHOR CONTRIBUTIONS**

MGK and LEH served as primary supervisors of this work. TJC, MGK, and ALS conceived and developed the analyses. TJC and WAB developed the OMOP2OBO algorithm with feedback from NAV and JMB. ALS and JMW helped develop documentation. RDB, AO, PBR, GH, JCD, DM, SJDD, and AW provided data for the evaluation. PNR, XAZ, MAH, NAM, SBT, EC, BDC, BM, JS, AYL, JHC, JR, JMW, ALS, JAF, TDB, NAV, KET, and CZ reviewed, evaluated or aided in pilot testing the mappings and/or assisted with the error analysis. CZ performed the *Clinical Utility* evaluation and JCD reviewed the results. TJC drafted the manuscript and all authors reviewed it, provided feedback, and approved the final version.

**COMPETING INTERESTS**

The authors declare no competing interests.

**FIGURE LEGENDS**

**Figure 1: Overview of the OMOP2OBO Algorithm.**

The OMOP2OBO algorithm[132] consists of three components: (1) **Process Input Data.** The algorithm takes as input a table of Observational Medical Outcomes Partnership (OMOP) concepts and a list of one or more OBO (Open Biological and Biomedical Ontology) Foundry ontologies. For both data types, the algorithm expects concept or class identifiers, source codes or database cross-references, labels, synonyms, and ancestor concepts or classes. (2) **Map OMOP Vocabulary Concepts to OBO Foundry Ontology Concepts.** This component is designed to automatically map or align OMOP concepts to OBO Foundry ontology concepts. The algorithm includes several different approaches, prioritizing those that result in high confident mappings. This component includes concept alignment and concept embedding.
(3) **Synthesize and Output Mapping Results.** The mapping results from the prior component are post-processed to include a mapping category and human-readable evidence. Post-processed mappings are serialized and able to be output to a variety of file types, like flat file, database dump, or RDF/XML file.

**Figure 2: OMOP2OBO Mapping Examples by OMOP Domain.**

This figure illustrates which OBO (Open Biological and Biomedical Ontology) Foundry ontologies were used for each OMOP (Observational Medical Outcomes Partnership) domain and provides example mappings. (**A**) OMOP conditions were mapped to HPO and Mondo. (**B**) OMOP drug ingredients were mapped to ChEBI, NCBITaxon, PRO, and VO. (**C**) OMOP measurements were mapped to ChEBI, CL, HPO, NCBITaxon, PRO, and Uberon.

Acronyms: UMLS (Unified Medical Language System); HP (Human Phenotype Ontology); MONDO (Monarch Disease Ontology); CHEBI (Chemical Entities of Biological Interest); NCBITaxon (National Center for Biotechnology Information Taxon Ontology); PR (Protein Ontology); VO (Vaccine Ontology); UBERON (Uber-Anatomy Ontology); CL (Cell Ontology).

**Figure 3: OMOP Concept Mapping Results by Domain, Concept Type, Mapping Category, and Ontology.**

The Sankey Diagram illustrates the mapping flow implemented by the OMOP2OBO algorithm beginning with OMOP (Observational Medical Outcomes Partnership) concepts from the Conditions, Drugs, and Measurements domains, which were grouped by Data Wave (i.e., whether or not the concept has been used at least once in clinical practice), and organized by mapping category. The flow lines in the diagram are weighted by the count of OMOP concepts from the Children's Hospital Colorado pediatric OMOP database.

Acronyms: OBO (Open Biological and Biomedical Ontology); HP (Human Phenotype Ontology); MONDO (Monarch Disease Ontology); CHEBI (Chemical Entities of Biological Interest); NCBITaxon (National Center for Biotechnology Information Taxon Ontology); PR (Protein Ontology); VO (Vaccine Ontology); UBERON (Uber-Anatomy Ontology); CL (Cell Ontology).

**Figure 4: Condition Concept Frequency of Use in Clinical Practice by Mapping Category and Ontology.**

This figure presents the frequency distributions of OMOP (Observational Medical Outcomes Partnership) condition concepts used at least once in clinical practice (log 10 scale) in the Children's Hospital Colorado pediatric OMOP database by mapping category and OBO (Open Biological and Biomedical Ontology) Foundry ontology. Center lines: median, boxes: first and third quartiles, whiskers: 1.5x interquartile range. The x-axis labels are numbers which correspond to the OMOP2OBO mapping categories: (1) Automatic One-to-One Concept; (2) Automatic One-to-One Ancestor (3) Automatic One-to-Many Concept; (4) Automatic One-to-Many Ancestor; (5) Cosine Similarity One-to-One Concept; (6) Manual One-to-One Concept; (7) Manual One-to-Many Concept; and (8) Unmapped.

Acronyms: HPO (Human Phenotype Ontology); Mondo (Monarch Disease Ontology).

**Figure 5: Drug Ingredient Concept Frequency of Use in Clinical Practice by Mapping Category and Ontology.**

This figure presents the frequency distributions of OMOP (Observational Medical Outcomes Partnership) drug exposure ingredient concepts used at least once in clinical practice (log 10 scale) in the Children's Hospital Colorado pediatric OMOP database by mapping category and OBO (Open Biological and Biomedical Ontology) Foundry ontology. Center lines: median, boxes: first and third quartiles, whiskers: 1.5x interquartile range. The x-axis labels are numbers which correspond to the OMOP2OBO mapping categories: (1) Automatic One-to-One Concept; (2) Automatic One-to-One Ancestor (3) Automatic One-to-Many Concept; (4) Automatic One-to-Many Ancestor; (5) Cosine Similarity One-to-One Concept; (6) Manual One-to-One Concept; (7) Manual One-to-Many Concept; and (8) Unmapped.

Acronyms: ChEBI (Chemical Entities of Biological Interest); NCBITaxon (National Center for Biotechnology Information Taxon Ontology); PRO (Protein Ontology); VO (Vaccine Ontology).

**Figure 6: Measurement Concept Frequency of Use in Clinical Practice by Mapping Category and Ontology.**

This figure presents the frequency distributions of OMOP (Observational Medical Outcomes Partnership) measurement concepts used at least once in clinical practice (log 10 scale) in the Children's Hospital Colorado pediatric OMOP database by mapping category and OBO (Open Biological and Biomedical Ontology) Foundry ontology. Center lines: median, boxes: first and third quartiles, whiskers: 1.5x interquartile range. The x-axis labels are numbers which correspond to the OMOP2OBO mapping categories: (1) Automatic One-to-One Concept; (2) Automatic One-to-One Ancestor (3) Automatic One-to-Many Concept; (4) Automatic One-to-Many Ancestor; (5) Cosine Similarity One-to-One Concept; (6) Manual One-to-One Concept; (7) Manual One-to-Many Concept; and (8) Unmapped.

Acronyms: HPO (Human Phenotype Ontology); Uberon (Uber-Anatomy Ontology); NCBITaxon (National Center for Biotechnology Information Taxon Ontology); PRO (Protein Ontology); ChEBI (Chemical Entities of Biological Interest); CL (Cell Ontology).

**Figure 7: OMOP2OBO - Concept Prevalence Coverage.**

This figure visualizes the coverage distributions of Observational Medical Outcomes Partnership (OMOP) concepts over their frequency of use in clinical practice (log 10 scale) within the Concept Prevalence Study data by domain ((**A**) Conditions; (**B**) Drugs; and (**C**) Measurements). The three modeled distributions include: concepts only found in the Concept Prevalence Study data (magenta), concepts only found in the OMOP2OBO mapping set (blue), and concepts found in both the Concept Prevalence Study data and the OMOP2OBO mapping set (yellow).

**Figure 8: OMOP2OBO - Concept Prevalence Coverage Error Analysis.**

This figure visualizes the distributions of Observational Medical Outcomes Partnership (OMOP) concepts missing from the OMOP2OBO mapping set over their frequency of use in clinical practice (log 10 scale) within the Concept Prevalence Study data by domain ((**A**) Conditions; (**B**) Drugs; and (**C**) Measurements). The three modeled error analysis distributions include: concepts recovered in a newer version of the OMOP common data model (CDM; magenta), concepts that were purposefully excluded, not yet mapped, or unable to be mapped by OMOP2OBO (blue), and concepts that were truly missing from the OMOP2OBO mapping set (yellow).

**Table 1:** Clinical Data and Ontologies Used for Input to OMOP2OBO Mapping Algorithm.

| OMOP CLINICAL VOCABULARIES | | | | |
|---|---|---|---|---|
| **Vocabulary** | **Concept Level** | **Concepts** | **Labels** | **Synonyms** |
| CONDITIONS | | | | |
| SNOMED-CT | *Standard Concepts Used In Practice* | | | |
| | Concept | 29129 | 29129 | 86630 |
| | Ancestor | 1421104 | 1389525 | N/A |
| | *Standard Concepts Not Used In Practice* | | | |
| | Concept | 80590 | 80590 | 194264 |
| | Ancestor | 3458072 | 3393343 | N/A |
| DRUG INGREDIENTS | | | | |
| RxNorm | *Standard Concepts Used In Practice* | | | |
| | Concept | 1693 | 1693 | 1865 |
| | Ancestor | 1697 | 1696 | N/A |
| | *Standard Concepts Not Used In Practice* | | | |
| | Concept | 10110 | 10110 | 11235 |
| | Ancestor | 10578 | 10578 | N/A |
| MEASUREMENTS | | | | |
| LOINC | *Standard Concepts Used In Practice* | | | |
| | Concept | 1606 | 1606 | 41917 |
| | Ancestor | 20784 | 21196 | N/A |
| | *Standard Concepts Not Used In Practice* | | | |
| | Concept | 2477 | 2477 | 73612 |
| | Ancestor | 23457 | 24306 | N/A |
| OBO FOUNDRY ONTOLOGIES | | | | |
| **Ontology** | **Classes** | **Labels** | **Synonyms** | **Cross-References** |
| ChEBI | 126169 | 126169 | 269798 | 231247 |
| CL | 2238 | 2238 | 2124 | 1376 |
| HPO | 15247 | 15247 | 19860 | 19569 |
| Mondo | 22288 | 22288 | 98181 | 159918 |
| NCBITaxon | 2241110 | 2241110 | 263571 | 18426 |
| PRO | 215624 | 215624 | 590190 | 195671 |
| Uberon | 13898 | 13898 | 36771 | 51322 |
| VO | 5789 | 5789 | 6 | 0 |

Acronyms: ChEBI (Chemical Entities of Biological Interest); CL (Cell Ontology); HPO (Human Phenotype Ontology); Mondo (Mondo Disease Ontology); NCBITaxon (National Center for Biotechnology Information Taxon Ontology); OMOP (Observational Medical Outcomes Partnership); PRO (Protein Ontology); Uberon (Uber-Anatomy Ontology); VO (Vaccine Ontology).

**Table 2:** Concept Prevalence Concepts Missing from the OMOP2OBO Mapping Set.

| Concept | Concept Label | ᵃAverage Concept Frequency | Study Site Count |
|---|---|---|---|
| *Conditions* | | | |
| 4091502 | Increased fluid intake | 100483.0 | 1 |
| 37311061 | COVID-19 | 93585.0 | 1 |
| 40443308 | Polycystic ovary syndrome | 62900.3 | 3 |
| 35615055 | Saddle embolus of pulmonary artery with acute cor pulmonale | 22324.4 | 10 |
| 36684319 | Adjustment disorder with mixed anxiety and depressed mood | 18453.0 | 1 |
| *Drug Ingredients* | | | |
| 37498625 | hepatitis A virus strain CR 326F antigen, inactivated | 175551.3 | 14 |
| 1510467 | erenumab | 60618.0 | 10 |
| 35200577 | fremanezumab | 15579.6 | 5 |
| 35200800 | galcanezumab | 11594.8 | 5 |
| 35201105 | baloxavir marboxil | 11366.7 | 3 |
| *Measurements* | | | |
| 3045980 | Pulse intensity of Unspecified artery palpation | 1219846862.0 | 1 |
| 3021716 | Penicillin G potassium [Mass] of Dose | 253609945.0 | 1 |
| 40760098 | Sodium [Moles/volume] in Saliva (oral fluid) | 246641311.0 | 1 |
| 3045820 | Cotinine/Creatinine [Mass Ratio] in Urine | 246063202.0 | 1 |
| 3008500 | Chloride [Moles/volume] in Saliva (oral fluid) | 234931483.0 | 1 |

ᵃThe average concept frequency was calculated as the frequency of each concept divided by the number of Concept Prevalence study sites with that concept by each clinical domain.

Concept labels were obtained from the Athena web application (https://athena.ohdsi.org/) on 12/29/2022.

**Supplementary Table 1:** Paper Acronyms and Concept Definitions.

| Term | Definition |
|---|---|
| *Acronyms* | |
| ACMG | American College of Medical Genetics and Genomics |
| AoU | All of Us Research Program |
| BoW | Bag-of-words |
| CDM | Common Data Model |
| CHCO | Children's Hospital of Colorado |
| ChEBI | Chemical Entities of Biological Interest |
| CL | Cell Ontology |
| CUI | Concept Unique Identifier |
| EHR | Electronic Health Record |
| eMERGE | Electronic Medical Records and Genomics |
| FBN1 | Fibrillin 1 |
| HPO | Human Phenotype Ontology |
| ICD | International Classification of Diseases |
| LOINC | Logical Observation Identifiers, Names and Codes |
| MEN1 | Menin 1 |
| Mondo | Mondo Disease Ontology |
| NCBITaxon | National Center for Biotechnology Information Organismal Taxonomy |
| NF2 | Moesin-Ezrin-Radixin Like (MERLIN) Tumor |
| OHDSI | Observational Health Data Sciences and Informatics |
| OBO | Open Biological and Biomedical Ontology |
| OMIM | Online Mendelian Inheritance in Man |
| OMOP | Observational Medical Outcomes Partnership |
| PEDSnet | National Pediatric Learning Health System |
| PheRS | Phenotype Risk Score |
| PRO | Protein Ontology |
| RET | Ret Proto-Oncogene |
| SDHAF2 | Succinate Dehydrogenase Complex Assembly Factor 2 |
| SDHB | Succinate Dehydrogenase Complex Subunit B |

| Term | Definition |
|---|---|
| SDHC | Succinate Dehydrogenase Complex Subunit C |
| SNOMED-CT | Systematized Nomenclature of Medicine -- Clinical Terms |
| TF-IDF | Term frequency-inverse document frequency |
| TGFBR1 | Transforming Growth Factor Beta Receptor 1 |
| TSC1 | Tuberous Sclerosis Complex Subunit 1 |
| TSC2 | Tuberous Sclerosis Complex Subunit 2 |
| Uberon | Uber-Anatomy Ontology |
| UMLS | Unified Medical Language System |
| VO | Vaccine Ontology |
| *Concepts* | |
| Concepts Used in Clinical Practice | Data Wave 1; All standard OMOP concepts used at least once in clinical practice |
| Concepts Not Used in Clinical Practice | Data Wave 2; All standard OMOP concepts not used in clinical practice |
| OMOP Standard Condition Occurrence Vocabulary | SnomedCT Release 20180131 |
| OMOP Standard Drug Exposure Ingredient Vocabulary | RxNorm Full 20180507 |
| OMOP Standard Measurement Vocabulary | LOINC 2.64 |
| OBO Foundry Ontologies mapped to OMOP Conditions | HPO, Mondo |
| OBO Foundry Ontologies mapped to OMOP Drug Ingredients | ChEBI, NCBITaxon, PRO, VO |
| OBO Foundry Ontologies mapped to OMOP Measurements | ChEBI, CL, HPO, NCBITaxon, PRO, Uberon |

**Supplementary Table 2:** OMOP2OBO Mapping Algorithm Resources.

| Resource | URL |
|---|---|
| *OMOP2OBO Resources* | |
| PyPI Package | https://pypi.org/project/omop2obo/ |
| GitHub Repository | https://github.com/callahantiff/OMOP2OBO |
| Project Wiki | https://github.com/callahantiff/OMOP2OBO/wiki |
| Mapping Dashboard | http://tiffanycallahan.com/OMOP2OBO_Dashboard/ |
| Zenodo Community | https://zenodo.org/communities/omop2obo |
| Condition Occurrence Mappings | https://doi.org/10.5281/zenodo.6774363 |
| Drug Exposure Ingredient Mappings | https://doi.org/10.5281/zenodo.6774401 |
| Measurement Mappings | https://doi.org/10.5281/zenodo.6774443 |
| Accuracy Evaluation | https://github.com/callahantiff/OMOP2OBO/wiki/Accuracy |
| Generalizability Evaluation | https://github.com/callahantiff/OMOP2OBO/wiki/Generalizability |
| *Mapping Resources* | |
| OMOP CDM V5.3 | https://ohdsi.github.io/CommonDataModel/cdm53.html |
| OHDSI Athena | https://athena.ohdsi.org/ |
| UMLS 2020AA Release Date | https://www.nlm.nih.gov/research/umls/licensedcontent/umlsarchives04.html#2020AA |
| LOINC2HPO Annotations | https://github.com/monarch-initiative/loinc2hpo/annotations.tsv |
| OHDSI Concept Prevalence Study | https://github.com/OHDSI/StudyProtocolSandbox/tree/master/ConceptPrevalence |
| *OBO Foundry Ontologies* | |
| ChEBI | http://purl.obolibrary.org/obo/chebi.owl |
| CL | http://purl.obolibrary.org/obo/cl.owl |
| HPO | http://purl.obolibrary.org/obo/hp.owl |
| Mondo | http://purl.obolibrary.org/obo/mondo.owl |
| NCBITaxon | http://purl.obolibrary.org/obo/ncbitaxon.owl |
| PRO | http://purl.obolibrary.org/obo/pr.owl |
| Uberon | http://purl.obolibrary.org/obo/uberon/ext.owl |
| VO | http://purl.obolibrary.org/obo/vo.owl |
| *Project and Analysis Notebooks* | |
| [a]OMOP2OBO | [b]OMOP2OBO/blob/master/omop2obo_notebook.ipynb |
| Mapping Analysis | [b]OMOP2OBO/blob/master/resources/analyses/omop2obo_manuscript_analyses.ipynb |
| Mapping Evaluation | [b]OMOP2OBO/blob/master/resources/analyses/omop2obo_mapping_validation.ipynb |

[a]This Jupyter Notebook serves the same purpose as the main.py script and provides users with a more interactive interface to use when running the algorithm.

[b]Primary OMOP2OBO Github: https://github.com/callahantiff/OMOP2OBO/.

Acronyms: ChEBI (Chemical Entities of Biological Interest); CL (Cell Ontology); HPO (Human Phenotype Ontology); Mondo (Mondo Disease Ontology); NCBITaxon (National Center for Biotechnology Information Taxon Ontology); PRO (Protein Ontology); Uberon (Uber-Anatomy Ontology); VO (Vaccine Ontology).

**Supplementary Table 3:** Clinical Data Used to Develop and Validate the OMOP2OBO Mappings.

| Data Source | Description | Use |
|---|---|---|
| CHCO OMOP Database | The CHCO pediatric OMOP database is a de-identified data repository that allows for the utilization of clinical pediatric information captured in electronic medical records. The database was created in October 2018, contains over 6 million patients, and is stored within University of Colorado Anschutz Medical Campus' Health Data Compass HIPAA Google Cloud-based infrastructure. The data conform to the structure defined by PEDSnet OMOP CDM v3.0, which is an adaptation of the OMOP CDM version 5.0. Use of these data was approved by the Colorado Multiple Institutional Review Board (#15-0445).<br><br>See GitHub[a] for more information: https://github.com/HealthDataCompass/CHCODeID | Mapping Development |
| OHDSI Concept Prevalence Data | The Concept Prevalence Study was conducted in order to examine patterns of OMOP standard concept use across several study sites within the OHDSI network. The data set includes OMOP standard concepts, OMOP domain, and record-level frequencies for each standard concept by study site. All study sites that contained data for standard OMOP condition, drug exposure ingredient, and measurement concepts were eligible for use in the current work (n=22 sites). These data were supplemented to include data from two additional academic medical centers. The 24 Study sites are listed below.<br><br>Study Sites: (1) Ajou University Database; (2) IQVIA US Ambulatory Electronic Medical Record; (3) IQVIA Longitudinal Patient Data Australia; (4) IQVIA Disease Analyzer France; (5) IQVIA Disease Analyzer Germany; (6) The Healthcare Cost and Utilization Project Nationwide Inpatient Sample; (7) IQVIA US Hospital Charge Data Master; (8) IBM MarketScan Commercial Database; (9) IBM MarketScan Multi-State Medicaid Database; (10) IBM MarketScan Medicare Supplemental Database; (11) Japan Medical Data Center database; (12) Medical Information Mart for Intensive Care III; (13) Korea National Health Insurance Service/National Sample Cohort; (14) Optum De-Identified Clinformatics Data-Mart-Database—Date of Death; (15) Optum De-Identified Clinformatics Data-Mart-Database—Socio-Economic Status; (16) Optum De-identified Electronic Health Record Dataset; (17) IQVIA US LRxDx Open Claims; (18) Premier Healthcare Database; (19) University of Southern California PScanner; (20) Stanford Medicine Research Data Repository; (21) Tufts Medical Center Database; (22) University of Colorado Anschutz Medical Campus Health Group; (23) Australian Electronic Practice-based Research Network; (24) Columbia University Medical Center Database.<br><br>See GitHub for more information: https://github.com/ohdsi-studies/ConceptPrevalence | Mapping Validation<br>*Generalization* |
| AoU Data | The National Institutes of Health's All of Us Research Program is an initiative tasked with gathering data from at least one million United States citizens with the goal of creating a diverse health resource to support biomedical research and precision medicine. The All of Us Research Hub contains data from over 630 sites on more than 528,000 participants. Data include electronic health records, biological and genetics samples, physical measurements and wearable data, and survey data. The All of Us Research Program would not be possible without the partnership of its participants. The current work utilized data from the version 6 build.<br><br>See the All of Us Research Hub for more information: https://www.researchallofus.org | Mapping Validation<br>*Clinical Utility* |

[a]This is a private repository, please contact the authors for access and to obtain additional information.

Acronyms: AoU (AllOfUs); CDM (common data model); CHCO (Children's Hospital Colorado); HIPAA (Health Insurance Portability and Accountability Act); OHDSI (Observational Health Data Sciences and Informatics); OMOP (Observational Medical Outcomes Partnership; PEDSnet (National Pediatric Learning Health System).

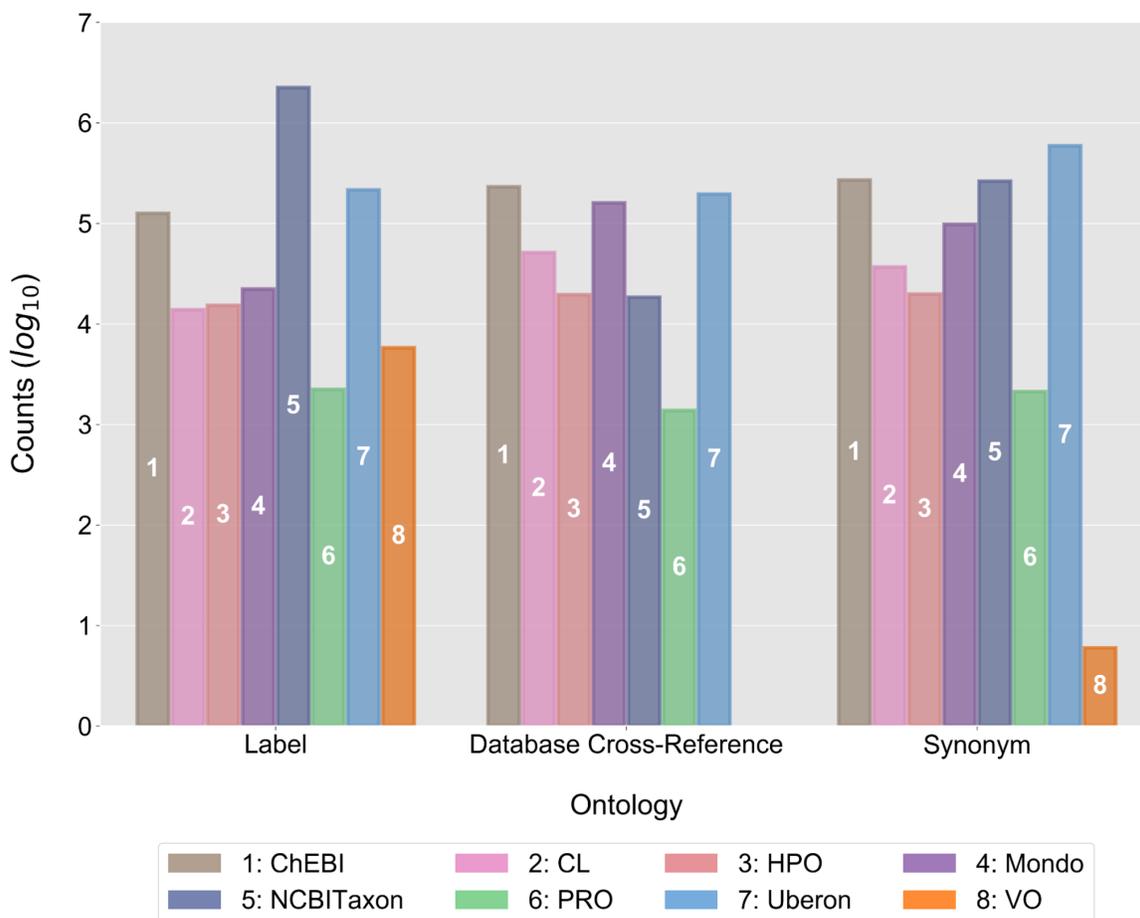

**Supplementary Figure 1: Available Mapping Metadata by OBO Foundry Ontology.**
This figure provides a visual illustration of the counts, in log 10 scale, of labels, database cross-references, and synonyms available for mapping by Open Biological and Biomedical Ontology (OBO) Foundry ontology. The labels on the bars are numbers which correspond to the ontologies: (1) ChEBI (Chemical Entities of Biological Interest); (2) CL (Cell Ontology); (3) HPO (Human Phenotype Ontology); (4) Mondo (Mondo Disease Ontology); (5) NCBITaxon (National Center for Biotechnology Information Taxon Ontology); (6) PRO (Protein Ontology); (7) Uberon (Uber-Anatomy Ontology); and (8) VO (Vaccine Ontology).

**Supplementary Table 4:** OMOP2OBO Mapping Categories.

| Mapping Category | Definition |
|---|---|
| Automatic One-to-One Concept | **Definition:** A one-to-one mapping that is automatically generated at the concept-level through exact string mappings to labels/synonyms or exact mappings between codes.<br><br>**Example:**<br>- `OMOP:22945` (Horizontal overbite)<br>- `HP:0011095` (Overjet)<br><br>This mapping was created through an exact string mapping on "overjet", which is the HP concept label and an OMOP concept synonym. This mapping is also supported through exact mappings between database cross-references to SNOMED-CT 70305005 and UMLS C0596028. |
| Automatic One-to-One Ancestor | **Definition:** A one-to-one mapping that is automatically generated for a concept's ancestor through exact string mappings to labels/synonyms or exact mappings between codes.<br><br>**Example:**<br>- `OMOP:22722` (Accessory salivary gland)<br>- `HP:0010286` (abnormal salivary gland morphology)<br><br>This mapping was created through exact mappings to one of the OMOP concept's ancestors on the database cross-references to SNOMED-CT 10890000 and UMLS C0036093. |
| Automatic One-to-Many Concept | **Definition:** A one-to-many mapping that is automatically generated at the concept-level through exact string mappings to labels/synonyms or exact mappings between codes. For release 1.0, one-to-many mappings indicate that one OMOP concept was mapped to one or more OBO Foundry ontology concepts.<br><br>**Example:**<br>- `OMOP:78854` (Osteopoikilosis)<br>- `MONDO:0001414` (Osteopoikilosis) AND `MONDO:0008157` (Duschke-Ollendorff Syndrome)<br><br>This mapping was created through 2 exact string mappings on "osteopoikilosis", which is a Mondo concept exact synonym and an OMOP concept label and synonym and "duschke-ollendorff syndrome", which is a Mondo concept exact synonym and label and an OMOP concept synonym. This mapping is also supported through exact mappings between database cross-references to SNOMED-CT 9147009. |
| Automatic One-to-Many Ancestor | **Definition:** A one-to-many mapping that is automatically generated for a concept's ancestor through exact string mappings to labels or synonyms or exact mappings between codes. For release 1.0, one-to-many mappings indicate that one OMOP concept was mapped to one or more OBO Foundry ontology concepts.<br><br>**Example:**<br>- `OMOP:74185` (Open fracture of cuboid bone of foot)<br>- `MONDO:0005315` (bone fracture) AND `MONDO:0044989` (foot disease)<br><br>This mapping was created through 3 exact string mappings on "fracture", "fracture of bone", and "disorder of foot", which are all Mondo exact synonyms and labels of the OMOP concept's ancestors. This mapping is also supported by exact mappings to one or more of the OMOP concept's ancestors on the database cross-references to SNOMED-CT 125605004 and 118932009. |

| Mapping Category | Definition |
|---|---|
| Manual One-to-One Concept | **Definition:** A one-to-one mapping that is manually generated at the concept-level and usually requires the use of external resources.<br>**Example:**<br>- `OMOP:4070954` (Mesiodens)<br>- `MONDO:0008533` (Teeth, supernumeracy)<br>This mapping was manually created through external evidence from a PubMed article, which stated "Mesiodens is a supernumerary tooth present in the midline between the two central incisors" (`PMID:21998774`). |
| Manual One-to-Many Concept | **Definition:** A one-to-many mapping that is manually generated at the concept-level and usually requires the use of external resources. For release 1.0, one-to-many mappings indicate that one OMOP concept was mapped to one or more OBO Foundry ontology concepts.<br>**Example:**<br>- `OMOP:439140` (Neonatal polycythemia)<br>- `HP:0003623` (Neonatal onset) AND `HP:0001901` (Polycythemia)<br>This mapping was created through an exact string mappings on "erythrocytosis", which is a HP concept exact synonym and a OMOP concept ancestor label. This mapping is also supported through exact mappings between database cross-references to SNOMED-CT 127062003 and UMLS C1527405 and C0032461. |
| Cosine Similarity One-to-One Concept | **Definition:** A one-to-one mapping that is automatically generated at the concept-level using cosine similarity scores. For release 1.0, the cosine similarity scores were applied to concept embeddings learned from a Bag-of-Words model with TF-IDF, which was applied to all available labels and synonyms at the concept- and ancestor-level.<br>**Example:**<br>- `OMOP:4147326` (Sore throat symptom)<br>- `HP:0033050` (Throat pain)<br>This mapping received a cosine similarity score of 0.66. |
| Unmapped | This concept is used when no suitable mapping is possible, for concepts which have not yet been mapped, and for concepts which are purposefully not mapped.<br>**Examples:**<br>*No Suitable Mondo Mapping*<br>- `OMOP:4235440` (Genetic alleles)<br>*Not Yet Mapped to HP or Mondo*<br>- `OMOP:4174055` (Athetoid paralysis)<br>*Purposefully Not Mapped to HP or Mondo*<br>- `OMOP:432499` (Mechanical complication due to coronary bypass graft) → *Complication*<br>- `OMOP:432498` (Burn of axilla) → *Injury*<br>- `OMOP:4056963` (Patient on self-medication) → *Finding* |

Acronyms: HP (Human Phenotype Ontology); Mondo (Mondo Disease Ontology); OMOP (Observational Medical Outcomes Partnership); PMID (PubMed Identifier); SNOMED-CT (Systematized Nomenclature of Medicine -- Clinical Terms); UMLS (Unified Medical Language System).

**Supplementary Table 5:** OMOP2OBO Condition Concept Mapping Results.

|  | HPO | | Mondo | |
|---|---|---|---|---|
| Concepts Used in Practice | Yes | No | Yes | No |
| *Mapping Category* | | | | |
| Automatic One-to-One Concept | 3601 | 1166 | 4836 | 4261 |
| Automatic One-to-One Ancestor | 3154 | 10440 | 5962 | 2949 |
| Automatic One-to-Many Concept | 125 | 25 | 632 | 253 |
| Automatic One-to-Many Ancestor | 1138 | 36947 | 4482 | 35742 |
| Cosine Similarity One-to-One Concept | 994 | 380 | 553 | 114 |
| Manual One-to-One Concept | 5119 | 0 | 755 | 0 |
| Manual One-to-Many Concept | 10328 | 0 | 2835 | 0 |
| **Total Mapped Concepts** | **24459** | **48958** | **20055** | **43319** |
| *Mapping Evidence* | | | | |
| Database Cross-References | 38473 | 279236 | 52430 | 339195 |
| Synonyms | 10169 | 42191 | 67381 | 85130 |
| Labels | 19343 | 97920 | 75795 | 113562 |
| Cosine Similarity | 11955 | 15825 | 12789 | 114 |
| Biocuration | 15447 | 0 | 3590 | 0 |
| **Total Mapping Evidence** | **95387** | **435172** | **211985** | **538001** |
| *Unmapped* | | | | |
| [a]None | 50 | 20771 | 84 | 5118 |
| Injury | 3323 | 10733 | 3323 | 10733 |
| Carrier Status | 23 | 0 | 22 | 0 |
| Complication | 906 | 128 | 906 | 128 |
| Finding | 368 | 0 | 4739 | 21292 |
| **Total Unmapped Concepts** | **4670** | **31632** | **9074** | **37271** |

The mapping category is constructed by combining the following elements: (1) the approach used to create it (i.e., "automatic", "manual", or "cosine similarity"), (2) cardinality (i.e., one-to-one or one-to-many), and (3) level (i.e., concept or ancestor).

[a]The unmapped "None" category for Concepts Not Used in Practice includes concepts that have not yet been mapped. For Concepts Used in Practice, "None" indicates concepts that were unable to be mapped to an Open Biological and Biomedical Ontology (OBO) Foundry ontology concept.

Acronyms: HPO (Human Phenotype); Mondo (Mondo Disease Ontology).

**Supplementary Table 6:** OMOP2OBO Drug Ingredient Concept Mapping Results.

| | ChEBI | | PRO | | VO | | NCBITaxon | |
|---|---|---|---|---|---|---|---|---|
| Concepts Used in Practice | Yes | No | Yes | No | Yes | No | Yes | No |
| *Mapping Category* | | | | | | | | |
| Automatic One-to-One Concept | 959 | 2192 | 1 | 42 | 90 | 18 | 20 | 135 |
| Automatic One-to-One Ancestor | 15 | 130 | 1 | 19 | 0 | 4 | 3 | 14 |
| Automatic One-to-Many Concept | 235 | 169 | 0 | 1 | 0 | 0 | 0 | 1 |
| Automatic One-to-Many Ancestor | 60 | 149 | 2 | 0 | 2 | 0 | 2 | 1 |
| Cosine Similarity One-to-One Concept | 31 | 78 | 8 | 10 | 3 | 14 | 136 | 4105 |
| Manual One-to-One Concept | 321 | 0 | 157 | 0 | 21 | 0 | 230 | 0 |
| Manual One-to-Many Concept | 72 | 0 | 8 | 0 | 2 | 0 | 14 | 0 |
| ***Total Mapped Concepts*** | **1693** | **2718** | **177** | **72** | **118** | **36** | **405** | **4256** |
| *Mapping Evidence* | | | | | | | | |
| Database Cross-References | 954 | 759 | 0 | 0 | 0 | 0 | 0 | 0 |
| Synonyms | 4565 | 7732 | 4 | 94 | 90 | 18 | 40 | 199 |
| Labels | 5573 | 9676 | 8 | 132 | 276 | 58 | 52 | 391 |
| Cosine Similarity | 1350 | 2562 | 9 | 54 | 96 | 32 | 160 | 4241 |
| Biocuration | 393 | 0 | 165 | 0 | 23 | 0 | 244 | 0 |
| ***Total Mapping Evidence*** | **12835** | **20729** | **186** | **280** | **485** | **108** | **496** | **4831** |
| *Unmapped* | | | | | | | | |
| [a]None | 0 | 7392 | 1516 | 10038 | 1575 | 10074 | 1288 | 5854 |
| ***Total Unmapped Concepts*** | **0** | **7392** | **1516** | **10038** | **1575** | **10074** | **1288** | **5854** |

The mapping category is constructed by combining the following elements: (1) the approach used to create it (i.e., "automatic", "manual", or "cosine similarity"), (2) cardinality (i.e., one-to-one or one-to-many), and (3) level (i.e., concept or ancestor).

[a]The unmapped "None" category for Concepts Not Used in Practice includes concepts that have not yet been mapped. For Concepts Used in Practice, "None" indicates concepts that were unable to be mapped to an Open Biological and Biomedical Ontology (OBO) Foundry ontology concept.

Acronyms: ChEBI (Chemical Entities of Biological Interest); PRO (Protein Ontology); VO (Vaccine Ontology); NCBITaxon (National Center for Biotechnology Information Taxon Ontology).

**Supplementary Table 7:** OMOP2OBO Measurement Concept Mapping Results.

|  | HPO | | Uberon | | NCBITaxon | | PRO | | ChEBI | | CL | |
|---|---|---|---|---|---|---|---|---|---|---|---|---|
| Concepts Used in Practice | Yes | No | Yes | No | Yes | No | Yes | No | Yes | No | Yes | No |
| *Mapping Category* | | | | | | | | | | | | |
| Automatic One-to-One Concept | 17 | 3 | 1793 | 3589 | 320 | 444 | 44 | 12 | 264 | 515 | 182 | 186 |
| Automatic One-to-One Ancestor | 23 | 20 | 592 | 593 | 181 | 351 | 9 | 6 | 1380 | 1924 | 14 | 0 |
| Automatic One-to-Many Concept | 0 | 0 | 10 | 0 | 0 | 0 | 0 | 0 | 0 | 0 | 46 | 24 |
| Automatic One-to-Many Ancestor | 0 | 0 | 2 | 0 | 0 | 0 | 0 | 0 | 29 | 3 | 3 | 0 |
| Cosine Similarity One-to-One Concept | 108 | 5 | 50 | 92 | 44 | 106 | 103 | 29 | 102 | 374 | 82 | 20 |
| Manual One-to-One Concept | 3902 | 6761 | 406 | 462 | 2300 | 4452 | 1267 | 2996 | 1377 | 2409 | 319 | 184 |
| Manual One-to-Many Concept | 37 | 12 | 1234 | 2065 | 5 | 454 | 149 | 189 | 337 | 1190 | 33 | 21 |
| **Total Mapped Concepts** | **4087** | **6801** | **4087** | **6801** | **2850** | **5807** | **1572** | **3232** | **3489** | **6415** | **679** | **435** |
| *Mapping Evidence* | | | | | | | | | | | | |
| Database Cross-References | 7 | 0 | 6 | 26 | 0 | 0 | 0 | 0 | 409 | 935 | 261 | 145 |
| Synonyms | 12 | 4 | 5232 | 8308 | 465 | 1627 | 73 | 24 | 2832 | 6166 | 486 | 414 |
| Labels | 28 | 24 | 1637 | 1242 | 307 | 458 | 29 | 14 | 3045 | 5712 | 296 | 227 |
| Cosine Similarity | 234 | 128 | 699 | 553 | 484 | 827 | 159 | 61 | 1482 | 2044 | 296 | 231 |
| Biocuration | 3939 | 6773 | 1640 | 2527 | 2305 | 4906 | 1416 | 3185 | 1714 | 3599 | 352 | 205 |
| **Total Mapping Evidence** | **4220** | **6929** | **9214** | **12656** | **3561** | **7818** | **1677** | **3284** | **9482** | **18456** | **1691** | **1222** |
| *Unmapped* | | | | | | | | | | | | |
| [a]None | 13 | 0 | 13 | 0 | 1250 | 994 | 2528 | 3569 | 611 | 386 | 3421 | 6366 |
| Not Mapped Test Type | 108 | 3 | 108 | 3 | 108 | 3 | 108 | 3 | 108 | 3 | 108 | 3 |
| Unspecified Sample | 217 | 40 | 217 | 40 | 217 | 40 | 217 | 40 | 217 | 40 | 217 | 40 |
| **Total Unmapped Concepts** | **338** | **43** | **338** | **43** | **1575** | **1037** | **2853** | **3612** | **936** | **429** | **3746** | **6409** |

The mapping category is constructed by combining the following elements: (1) the approach used to create it (i.e., "automatic", "manual", or "cosine similarity"), (2) cardinality (i.e., one-to-one or one-to-many), and (3) level (i.e., concept or ancestor).

[a]The unmapped "None" category for Concepts Not Used in Practice includes concepts that have not yet been mapped. For Concepts Used in Practice, "None" indicates concepts that were unable to be mapped to an Open Biological and Biomedical Ontology (OBO) Foundry ontology concept.

Acronyms: HPO (Human Phenotype Ontology); Uberon (Uber-Anatomy Ontology); NCBITaxon (National Center for Biotechnology Information Taxon Ontology); PRO (Protein Ontology); ChEBI (Chemical Entities of Biological Interest); CL (Cell Ontology).

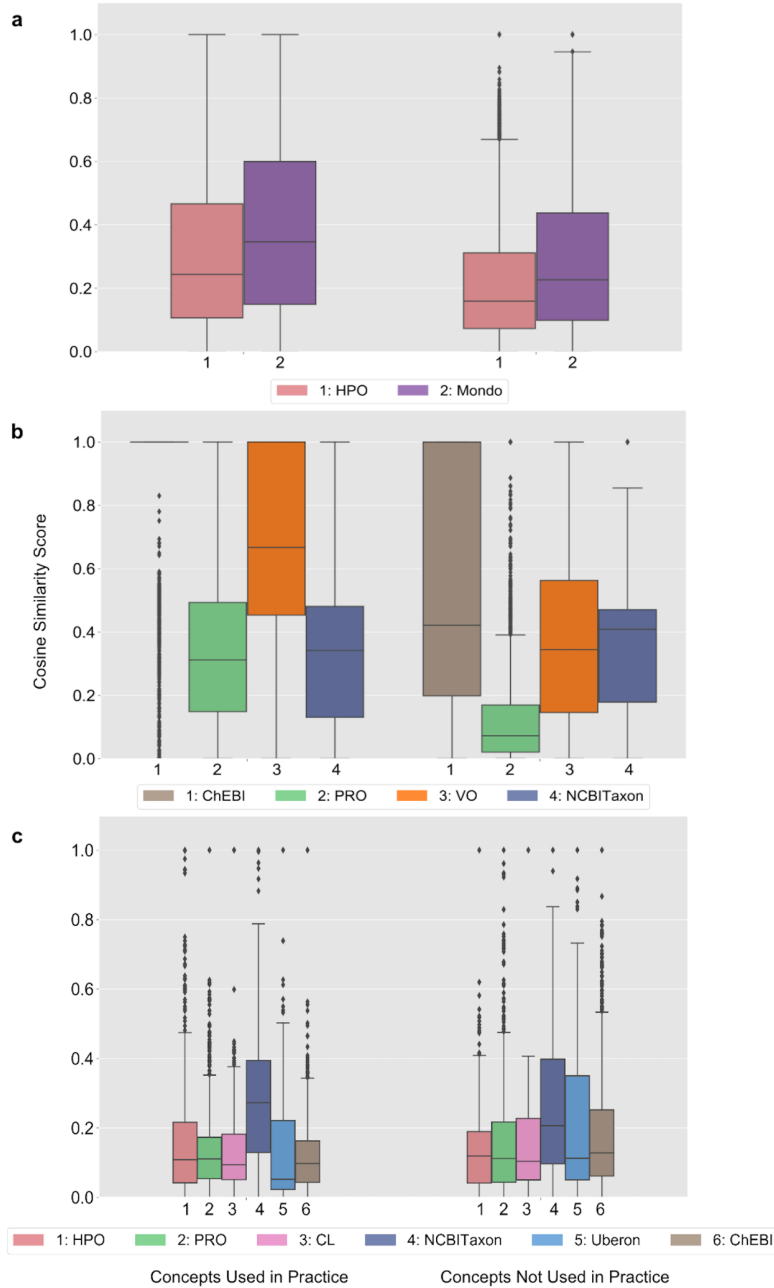

**Supplementary Figure 2: Concept Similarity Scores by OMOP Domain and OBO Foundry Ontology.**
The figure presents the distribution of cosine similarity scores by Open Biological and Biomedical Ontology (OBO) Foundry ontology and data wave (Concepts Used in Practice [all concepts associated with at least one patient and/or visit in the Children's Hospital of Colorado OMOP Database] and Concepts Not Used in Practice [all concepts not used in clinical practice]) for three Observational Medical Outcomes Partnership (OMOP) domains: (**A**) Conditions, (**B**) Drugs, and (**C**) Measurements. Center lines: median, boxes: first and third quartiles, whiskers: 1.5x interquartile range. The x-axis labels are numbers which correspond to the ontologies within each domain from top to bottom: Conditions (1: HPO, 2: Mondo); Drug Ingredients (1: ChEBI, 2: PRO, 3: VO, 4: NCBITaxon); and Measurements (1: HPO, 2: PRO, 3: CL, 4: NCBITaxon, 5: Uberon, 6: ChEBI).

Acronyms: HPO (Human Phenotype Ontology); Mondo (Monarch Disease Ontology); ChEBI (Chemical Entities of Biological Interest); PRO (Protein Ontology); VO (Vaccine Ontology); NCBITaxon (National Center for Biotechnology Information Taxon Ontology); CL (Cell Ontology); Uberon (Uber-Anatomy Ontology).

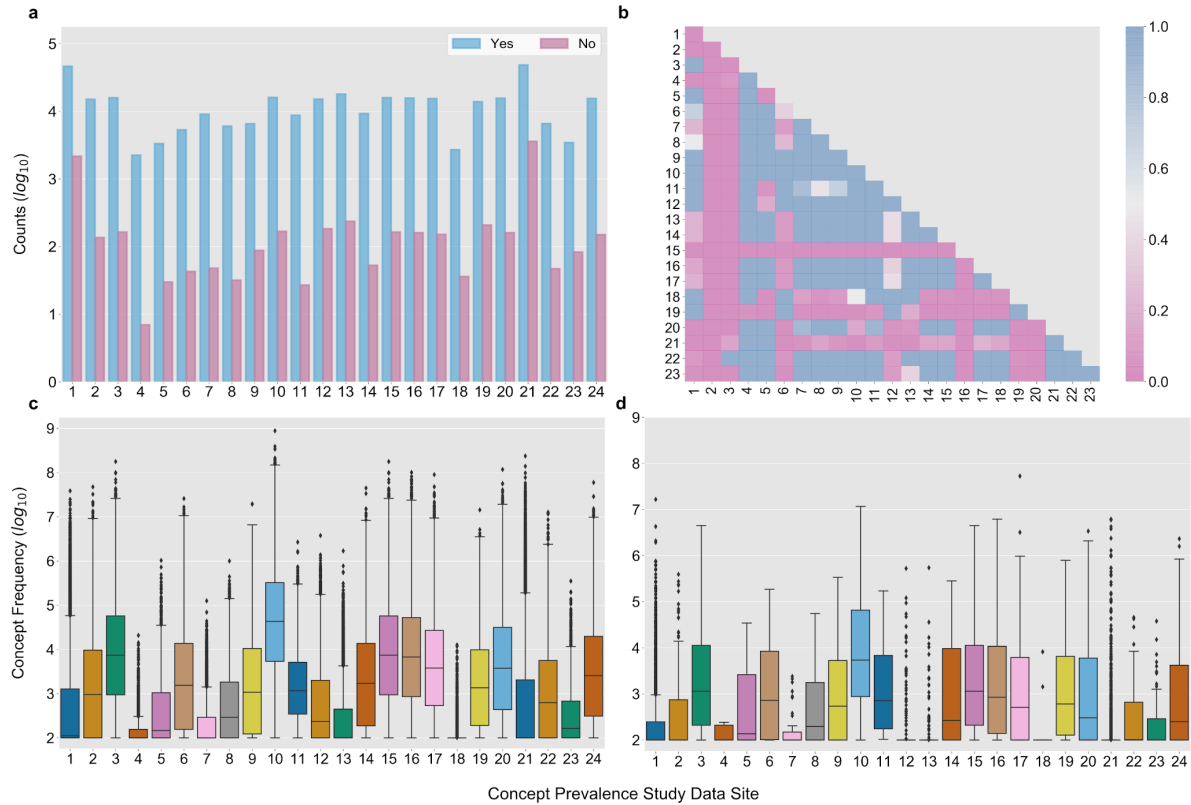

**Supplementary Figure 3: Overview of the OMOP2OBO Condition Concepts in the OHDSI Concept Prevalence Data by Coverage Status.**
(**A**) This figure presents the counts of OMOP (Observational Medical Outcomes Partnership) condition concepts (log 10 scale) in the Concept Prevalence Study data by site (1-24) and whether or not they are covered by the OMOP2OBO mapping set ("Yes"/"No"). (**B**) This figure visualizes the results of conducting a Chi-square test of independence with Yate's correction to assess differences in the proportions of OMOP condition concepts covered by the OMOP2OBO mapping set across the Concept Prevalence Study data sites. The figure presents a heatmap to visualize Bonferroni adjusted p-values for post-hoc tests which confirmed that 32% of the pairwise site comparisons had significantly different coverage of the OMOP2OBO mapping sets (ps<0.001 for all significant comparisons). (**C**) This figure presents the frequency distributions of OMOP condition concepts covered by the OMOP2OBO mapping set (log 10 scale) in the Concept Prevalence Study data by site. (**D**) This figure presents the frequency distributions of OMOP condition concepts not covered by the OMOP2OBO mapping set (log 10 scale) in the Concept Prevalence Study data by site. Figures **C**-**D**: Center lines (median), boxes (first and third quartiles), whiskers (1.5x interquartile range). The x-axis labels are numbers which correspond to the Concept Prevalence Study site index: (1) Ajou University Database; (2) IQVIA US Ambulatory Electronic Medical Record; (3) IQVIA Longitudinal Patient Data Australia; (4) IQVIA Disease Analyzer France; (5) IQVIA Disease Analyzer Germany; (6) The Healthcare Cost and Utilization Project Nationwide Inpatient Sample; (7) IQVIA US Hospital Charge Data Master; (8) IBM MarketScan Commercial Database; (9) IBM MarketScan Multi-State Medicaid Database; (10) IBM MarketScan Medicare Supplemental Database; (11) Japan Medical Data Center database; (12) Medical Information Mart for Intensive Care III; (13) Korea National Health Insurance Service/National Sample Cohort; (14) Optum De-Identified Clinformatics Data-Mart-Database—Date of Death; (15) Optum De-Identified Clinformatics Data-Mart-Database—Socio-Economic Status; (16) Optum De-identified Electronic Health Record Dataset; (17) IQVIA US LRxDx Open Claims; (18) Premier Healthcare Database; (19) University of Southern California PScanner; (20) Stanford Medicine Research Data Repository; (21) Tufts Medical Center Database; (22) University of Colorado Anschutz Medical Campus Health Group; (23) Australian Electronic Practice-based Research Network; (24) Columbia University Medical Center Database.

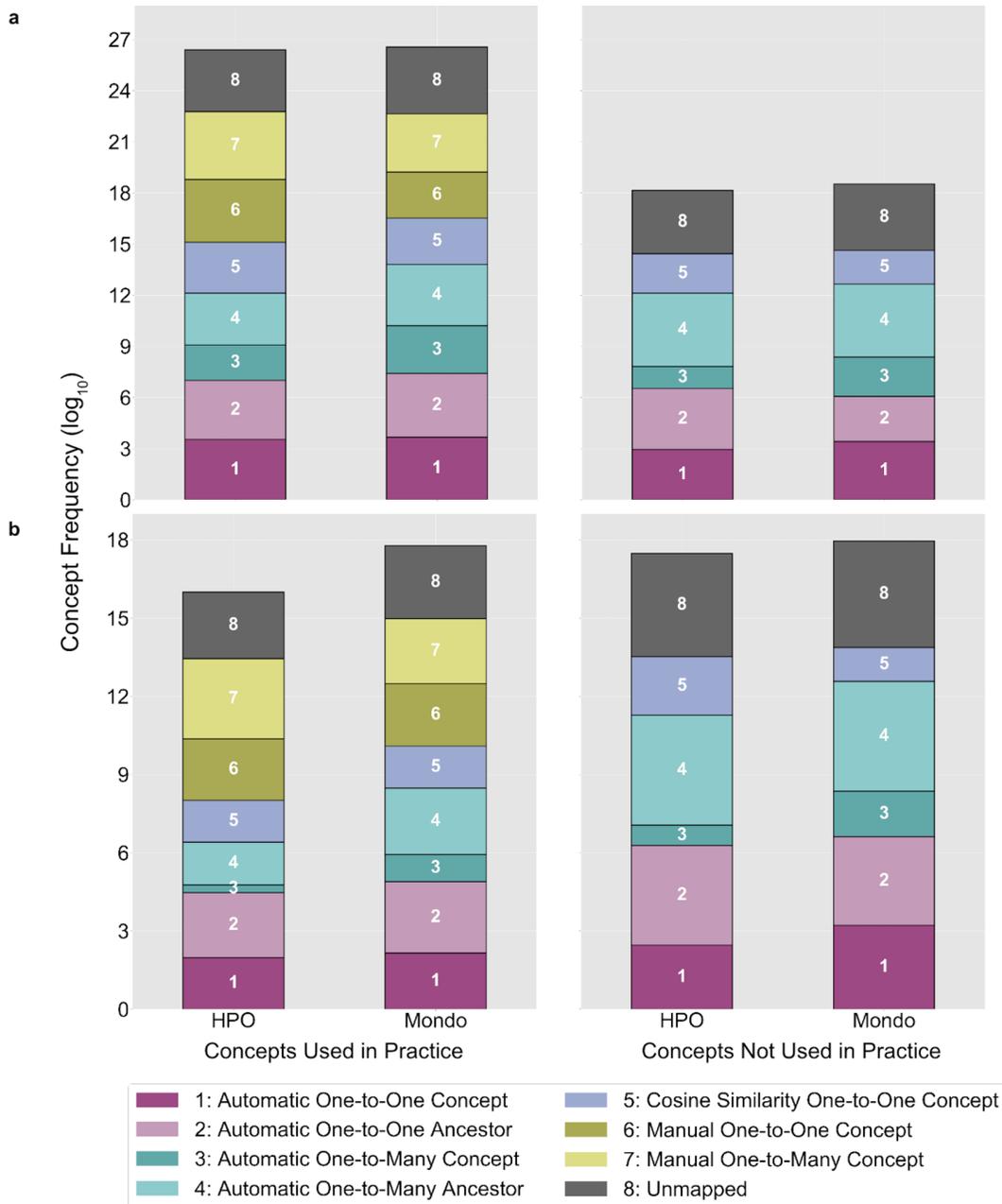

**Supplementary Figure 4: Frequency of OMOP2OBO Condition Concepts in the OHDSI Concept Prevalence Data by OBO Foundry Ontology and Data Wave.**
(**A**) This figure visualizes the count of Observational Medical Outcomes Partnership (OMOP) condition concepts (log 10 scale) in the OMOP2OBO mapping set that overlapped with Concept Prevalence Study by Open Biological and Biomedical Ontology (OBO) Foundry ontology and (Concepts Used in Practice [all concepts associated with at least one patient and/or visit in the Children's Hospital of Colorado OMOP Database] and Concepts Not Used in Practice [all concepts not used in clinical practice]). (**B**) This figure visualizes the count of OMOP condition concepts (log 10 scale) in the OMOP2OBO mapping set condition concepts that were not present in the Concept Prevalence Study data by OBO Foundry ontology and data wave. The labels on the bars are numbers which correspond to the OMOP2OBO mapping categories: (1) Automatic One-to-One Concept; (2) Automatic One-to-One Ancestor (3) Automatic One-to-Many Concept; (4) Automatic One-to-Many Ancestor; (5) Cosine Similarity One-to-One Concept; (6) Manual One-to-One Concept; (7) Manual One-to-Many Concept; and (8) Unmapped.

Acronyms: HPO (Human Phenotype Ontology); Mondo (Monarch Disease Ontology).

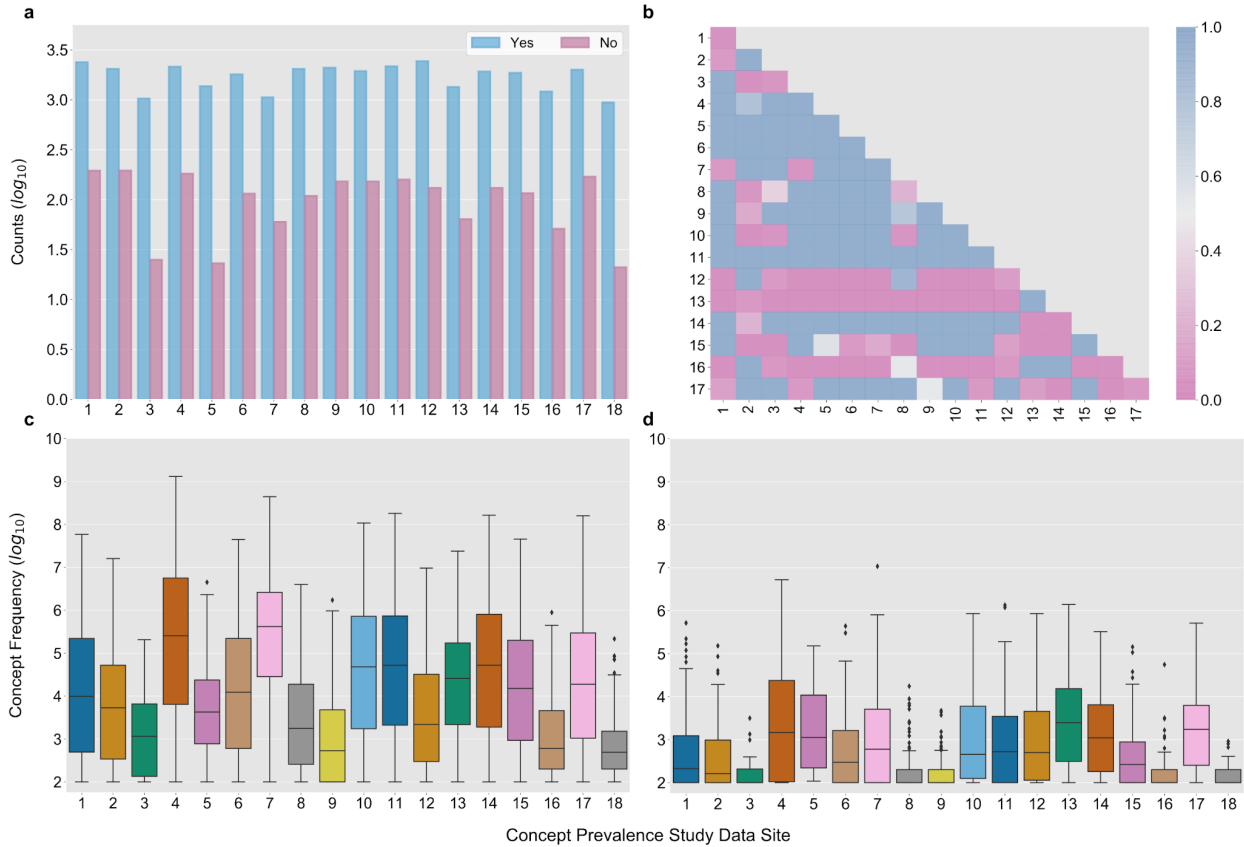

**Supplementary Figure 5: Overview of the OMOP2OBO Drug Ingredient Concepts in the OHDSI Concept Prevalence Data by Coverage Status.**
(**A**) This figure presents the counts of OMOP (Observational Medical Outcomes Partnership) drug ingredient concepts (log 10 scale) in the Concept Prevalence Study data by site (1-18) and whether or not they are covered by the OMOP2OBO mapping set ("Yes"/"No"). (**B**) This figure visualizes the results of conducting a Chi-square test of independence with Yate's correction to assess differences in the proportions of OMOP drug ingredient concepts covered by the OMOP2OBO mapping set across the Concept Prevalence Study data sites. The figure presents a heatmap to visualize Bonferroni adjusted p-values for post-hoc tests which confirmed that 22% of the pairwise site comparisons had significantly different coverage of the OMOP2OBO mapping sets (ps<0.001 for all significant comparisons). (**C**) This figure presents the frequency distributions of OMOP drug ingredient concepts covered by the OMOP2OBO mapping set (log 10 scale) in the Concept Prevalence Study data by site. (**D**) This figure presents the frequency distributions of OMOP drug ingredient concepts not covered by the OMOP2OBO mapping set (log 10 scale) in the Concept Prevalence Study data by site. Figures **C**-**D**: Center lines (median), boxes (first and third quartiles), whiskers (1.5x interquartile range). The x-axis labels are numbers which correspond to the Concept Prevalence Study site index: (1) IQVIA US Ambulatory Electronic Medical Record; (2) IQVIA Longitudinal Patient Data Australia; (3) IQVIA Disease Analyzer Germany; (4) IQVIA US Hospital Charge Data Master; (5) IBM MarketScan Commercial Database; (6) IBM MarketScan Multi-State Medicaid Database; (7) IBM MarketScan Medicare Supplemental Database; (8) Japan Medical Data Center database; (9) Optum De-Identified Clinformatics Data-Mart-Database—Socio-Economic Status; (10) Optum De-identified Electronic Health Record Dataset; (11) Optum De-identified Electronic Health Record Dataset; (12) Premier Healthcare Database; (13) University of Southern California PScanner; (14) Stanford Medicine Research Data Repository; (15) Tufts Medical Center Database; (16) University of Colorado Anschutz Medical Campus Health Group; (17) Australian Electronic Practice-based Research Network; (18) Columbia University Medical Center Database.

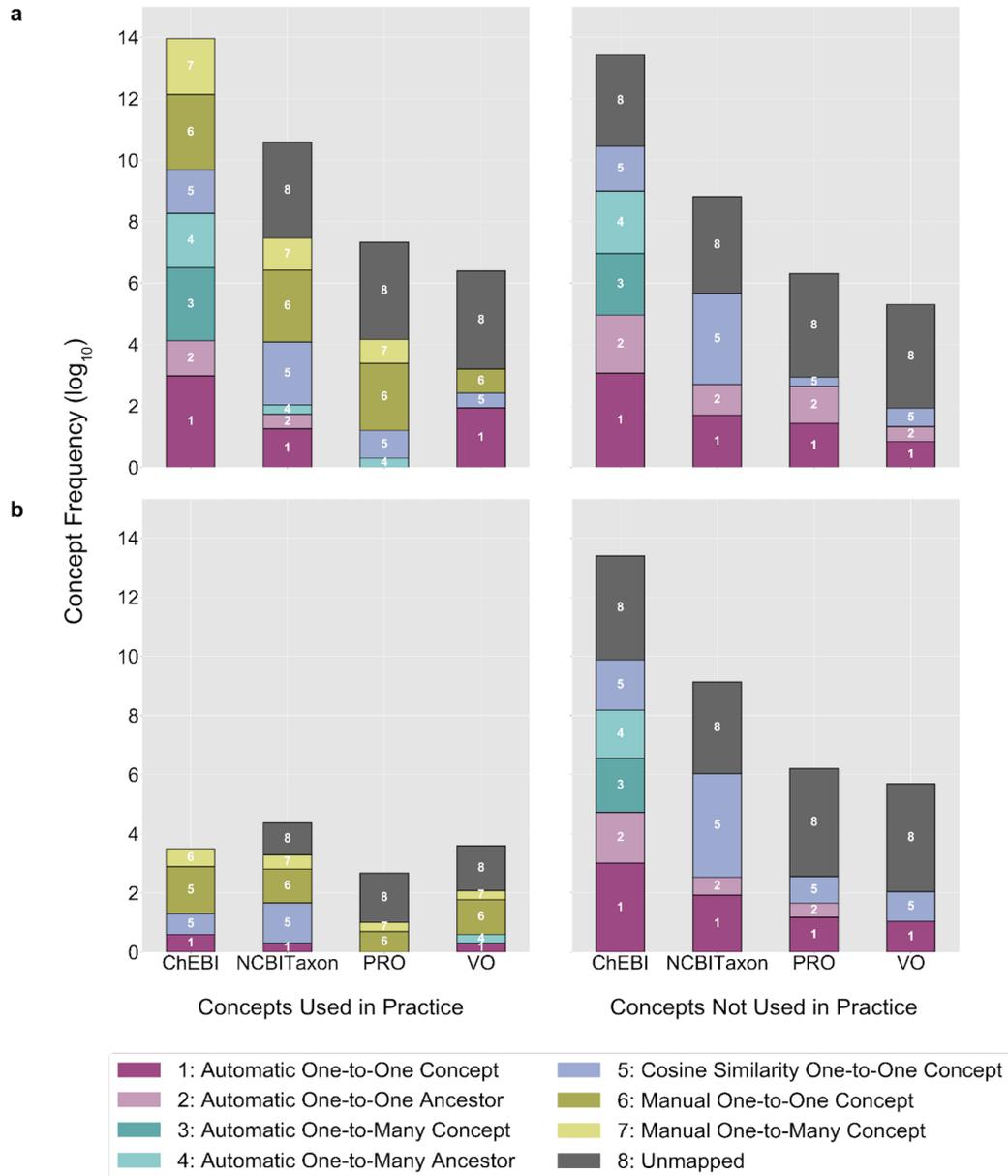

**Supplementary Figure 6: Frequency of OMOP2OBO Drug Ingredient Concepts in the OHDSI Concept Prevalence Data by OBO Foundry Ontology and Data Wave.**
(**A**) This figure visualizes the count of Observational Medical Outcomes Partnership (OMOP) drug ingredient concepts (log 10 scale) in the OMOP2OBO mapping set that overlapped with concepts in the Concept Prevalence Study by Open Biological and Biomedical Ontology (OBO) Foundry ontology and data wave (Concepts Used in Practice [all concepts associated with at least one patient and/or visit in the Children's Hospital of Colorado OMOP Database] and Concepts Not Used in Practice [all standard OMOP concepts not used in clinical practice]). (**B**) This figure visualizes the count of OMOP drug ingredient concepts (log 10 scale) in the OMOP2OBO mapping set that were not present in the Concept Prevalence Study data by OBO Foundry ontology and data wave. The labels on the bars are numbers which correspond to the OMOP2OBO mapping categories: (1) Automatic One-to-One Concept; (2) Automatic One-to-One Ancestor (3) Automatic One-to-Many Concept; (4) Automatic One-to-Many Ancestor; (5) Cosine Similarity One-to-One Concept; (6) Manual One-to-One Concept; (7) Manual One-to-Many Concept; and (8) Unmapped.

Acronyms: ChEBI (Chemical Entities of Biological Interest); NCBITaxon (National Center for Biotechnology Information Taxon Ontology); PRO (Protein Ontology); VO (Vaccine Ontology).

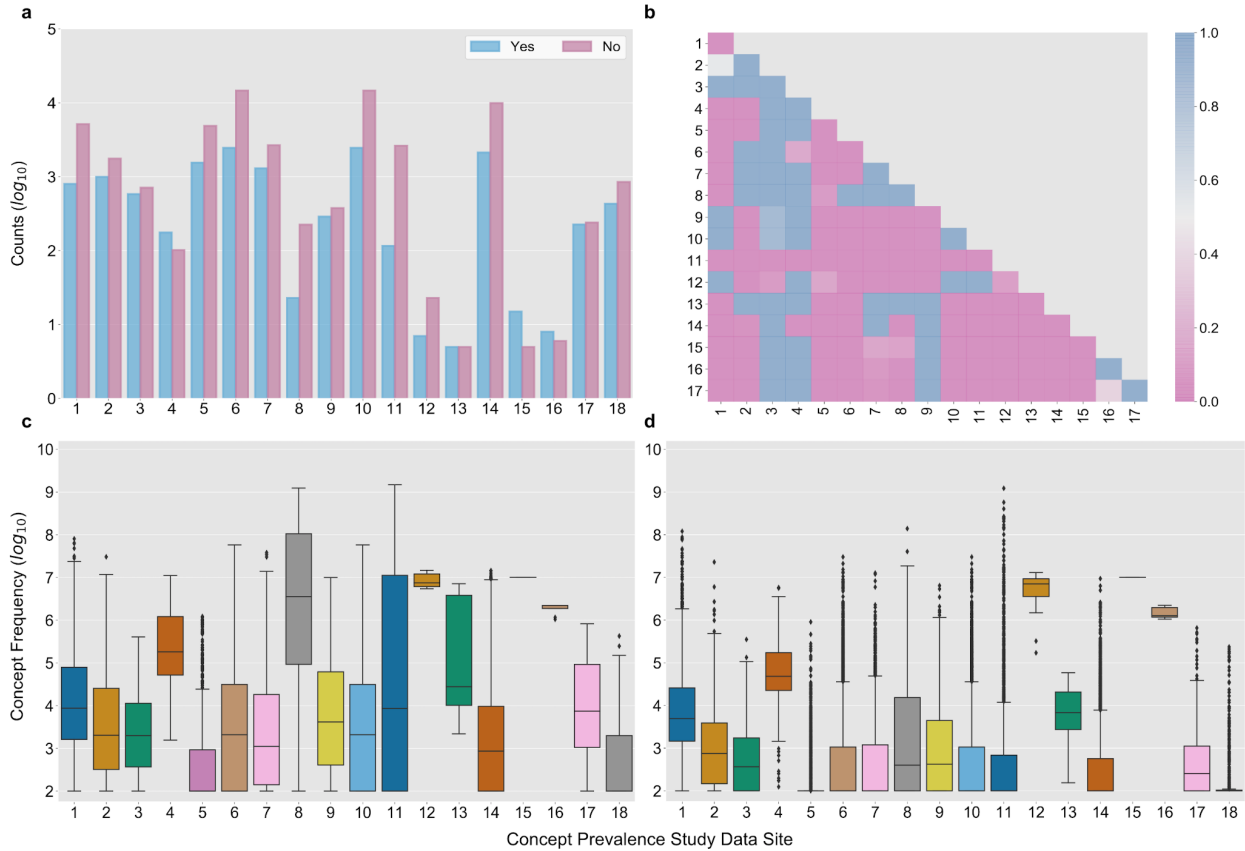

**Supplementary Figure 7: Overview of the OMOP2OBO Measurement Concepts in the OHDSI Concept Prevalence Data by Coverage Status.**
(**A**) This figure presents the counts of OMOP (Observational Medical Outcomes Partnership) measurement concepts (log 10 scale) in the Concept Prevalence Study data by site (1-18) and whether or not they are covered by the OMOP2OBO mapping set ("Yes"/"No"). (**B**) This figure visualizes the results of conducting a Chi-square test of independence with Yate's correction to assess differences in the proportions of OMOP measurement concepts covered by the OMOP2OBO mapping set across the Concept Prevalence Study data sites. The figure presents a heatmap to visualize Bonferroni adjusted p-values for post-hoc tests which confirmed that 56% of the pairwise site comparisons had significantly different coverage of the OMOP2OBO mapping sets (ps<0.001 for all significant comparisons). (**C**) This figure presents the frequency distributions of OMOP measurement concepts covered by the OMOP2OBO mapping set (log 10 scale) in the Concept Prevalence Study data by site. (**D**) This figure presents the frequency distributions of OMOP measurement concepts not covered by the OMOP2OBO mapping set (log 10 scale) in the Concept Prevalence Study data by site. Figures **C**-**D**: Center lines (median), boxes (first and third quartiles), whiskers (1.5x interquartile range). The x-axis labels are numbers which correspond to the Concept Prevalence Study site index: (1) IQVIA US Ambulatory Electronic Medical Record; (2) IQVIA Longitudinal Patient Data Australia; (3) IQVIA Disease Analyzer France; (4) IQVIA Disease Analyzer Germany; (5) IBM MarketScan Commercial Database; (6) IBM MarketScan Medicare Supplemental Database; (7) Japan Medical Data Center database; (8) Medical Information Mart for Intensive Care III; (9) Korea National Health Insurance Service/National Sample Cohort; (10) Optum De-Identified Clinformatics Data-Mart-Database—Date of Death; (11) Optum De-Identified Clinformatics Data-Mart-Database—Socio-Economic Status; (12) Optum De-identified Electronic Health Record Dataset; (13) Premier Healthcare Database; (14) University of Southern California PScanner; (15) Stanford Medicine Research Data Repository; (16) University of Colorado Anschutz Medical Campus Health Group; (17) Australian Electronic Practice-based Research Network; (18) Columbia University Medical Center Database.

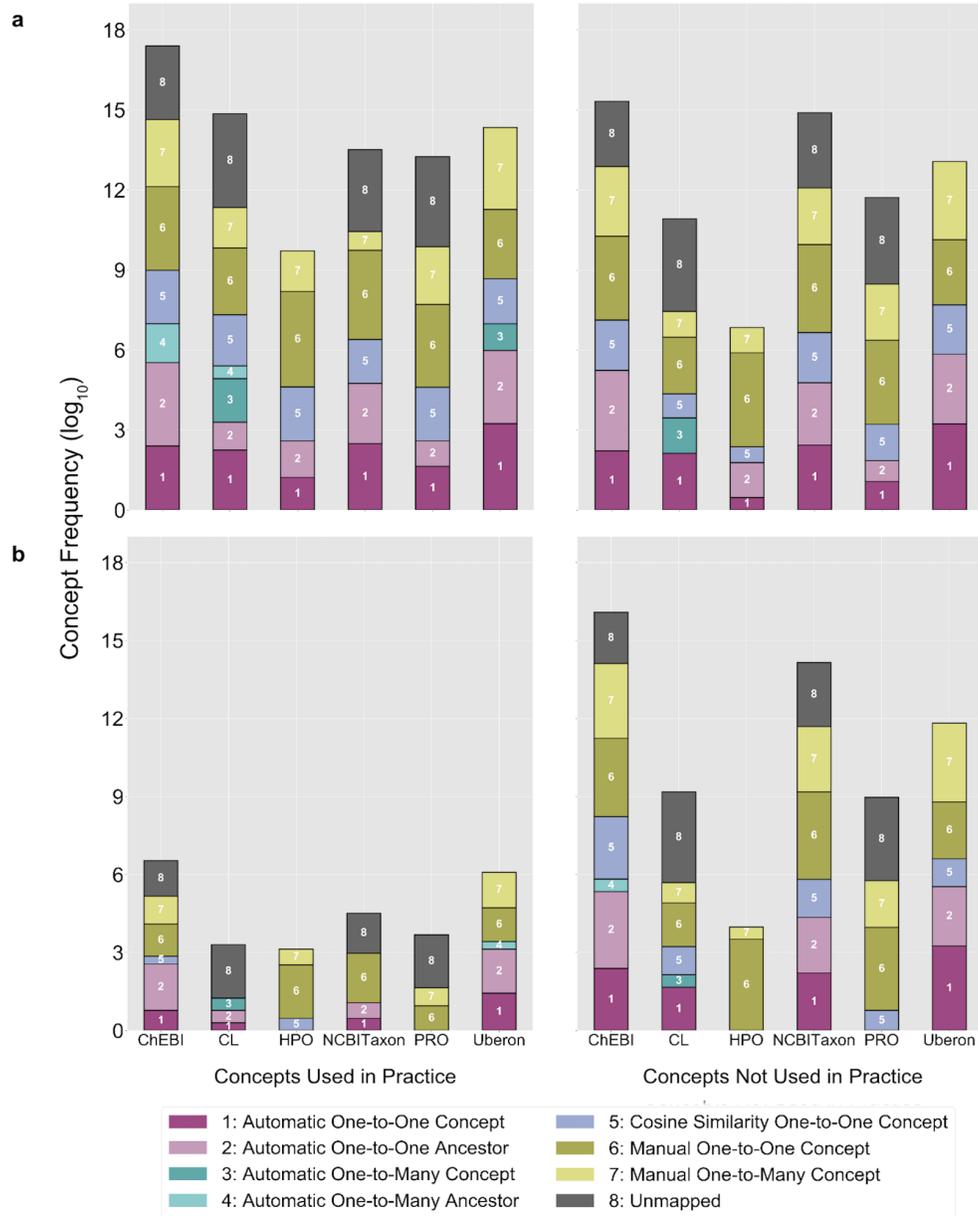

**Supplementary Figure 8: Frequency of OMOP2OBO Measurement Concepts in the OHDSI Concept Prevalence Data by OBO Foundry Ontology and Data Wave.**

(**A**) This figure visualizes the count of Observational Medical Outcomes Partnership (OMOP) measurement concepts (log 10 scale) in the OMOP2OBO mapping set that overlapped with concepts in the Concept Prevalence Study by Open Biological and Biomedical Ontology (OBO) Foundry ontology and data wave (Concepts Used in Practice [all concepts associated with at least one patient and/or visit in the Children's Hospital of Colorado OMOP Database] and Concepts Not Used in Practice [all concepts not used in clinical practice]). (**B**) This figure visualizes the count of OMOP measurement concepts (log 10 scale) in the OMOP2OBO mapping set that were not present in the Concept Prevalence Study data by OBO Foundry ontology and data wave. The labels on the bars are numbers which correspond to the OMOP2OBO mapping categories: (1) Automatic One-to-One Concept; (2) Automatic One-to-One Ancestor; (3) Automatic One-to-Many Concept; (4) Automatic One-to-Many Ancestor; (5) Cosine Similarity One-to-One Concept; (6) Manual One-to-One Concept; (7) Manual One-to-Many Concept; and (8) Unmapped.

Acronyms: ChEBI (Chemical Entities of Biological Interest); CL (Cell Ontology); HPO (Human Phenotype Ontology); NCBITaxon (National Center for Biotechnology Information Taxon Ontology); PRO (Protein Ontology); Uberon (Uber-Anatomy Ontology).

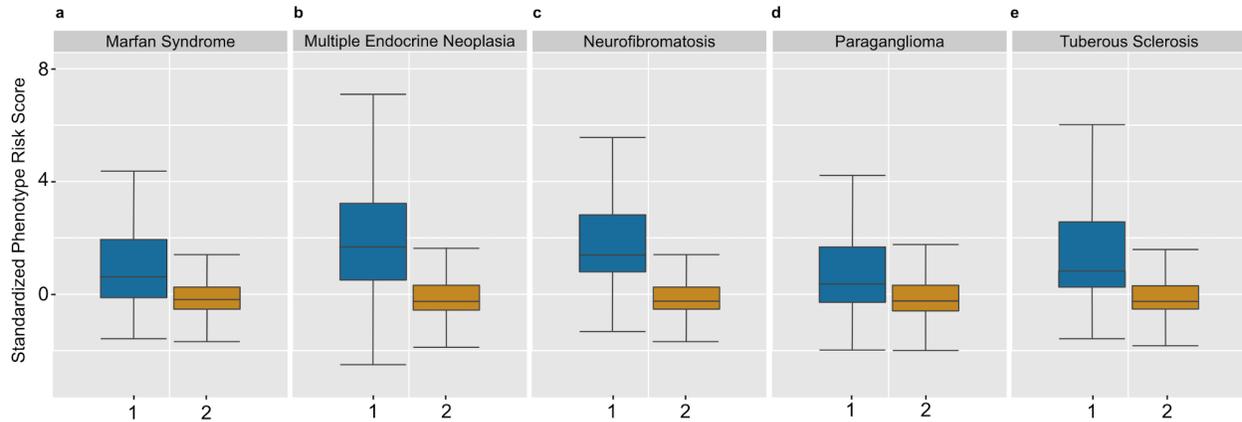

**Supplementary Figure 9: Standardized Phenotype Risk Scores (PheRS) by Disease for Cases and Controls.**
The Phenotype Risk Score (PheRS) is a measure used to identify patients with phenotypic features that are clinically similar to Online Mendelian Inheritance in Man (OMIM) Mendelian profiles but who lack formal diagnosis and has demonstrated utility for identifying underdiagnosed rare disease patients using only electronic health record data. The standardized PheRS was applied to five diseases (Figures **A-E**) known to be caused by pathogenic genetic mutations in 11 American College of Medical Genetics and Genomics secondary finding genes (listed by relevant disease below). In this figure, boxplots of the PheRS are used to illustrate differences between cases and controls for each of the five diseases using data from the All of Us Research Program. To determine if the PheRSs were significantly higher for cases than controls, one-sided Wilcoxon rank sum tests were performed for each disease. Results confirmed that cases had significantly higher PheRS than controls for all examined diseases ($p<0.001$ across all diseases), which included: (**A**) Marfan syndrome (*FBN1*, *TGFBR1*); (**B**) multiple endocrine neoplasia related to (*MEN1*, *RET*); (**C**) neurofibromatosis (*NF2*); (**D**) paragangliomas (related to succinate dehydrogenase genes: *SDHAF2*, *SDHB*, *SDHC*, *SDHD*); and (**E**) tuberous sclerosis complex (*TSC1*, *TSC2*). Center lines: median, boxes: first and third quartiles, whiskers: 1.5x interquartile range. The x-axis labels are numbers which correspond to (1) control (blue) and (2) case (yellow) patients.

**Supplementary Table 8:** Descriptive Statistics by Disease for Cases and Controls.

|  | Marfan Syndrome | Multiple Endocrine Neoplasia | Neurofibromatosis | Paraganglioma | Tuberous Sclerosis |
|---|---|---|---|---|---|
| *Cases* | | | | | |
| Patient Count | 131 | 86 | 255 | 105 | 38 |
| *Standardized PheRS[a]* | | | | | |
| Mean | 1.136 | 2.147 | 1.968 | 1.072 | 1.317 |
| Median | 0.616 | 1.673 | 1.381 | 0.378 | 0.824 |
| Standard Deviation | 2.02 | 2.375 | 1.981 | 2.308 | 1.811 |
| Range (min, max) | -3.326, 11.521 | -2.512, 11.402 | -1.305, 10.767 | -1.970, 10.249 | -1.578, 6.003 |
| *Controls* | | | | | |
| Patient Count | 63,086 | 72,150 | 65,256 | 68,552 | 58,555 |
| *Standardized PheRS[a]* | | | | | |
| Mean | -0.013 | -0.004 | -0.006 | -0.002 | -0.009 |
| Median | -0.186 | -0.245 | -0.234 | -0.239 | -0.264 |
| Standard Deviation | 0.949 | 0.996 | 0.993 | 1.001 | 0.989 |
| Range (min, max) | -12.476, 7.366 | -12.305, 11.213 | -9.393, 13.595 | -9.919, 13.539 | -10.544, 23.098 |

[a]The standardized PheRS is derived by subtracting the normalized raw scores by the mean and dividing by the standard deviation.
Acronyms: PheRS (Phenotype Risk Score)